\documentclass[reprint,floatfix,numerical,longbibliography,superscriptaddress]{revtex4-1}
\usepackage{amsmath}
\usepackage{hyperref}
\usepackage{graphicx}
\usepackage{siunitx}
\usepackage{color}

\newcommand{\gdot}{\dot{\gamma}}

\definecolor{grey}{gray}{0.5}

\hypersetup{
  pdftitle={Unsteady flow of very dense non-Brownian suspensions},
  pdfauthor={Hermes et al.},
  colorlinks=true,
  citecolor=black,
  linkcolor=black,
  urlcolor=blue
}

\begin{document}

\title{Unsteady flow and particle migration in dense, non-Brownian suspensions}
\author{Michiel Hermes}
\affiliation{School of Physics and Astronomy, The University of Edinburgh, King's Buildings, Peter Guthrie Tait Road, Edinburgh, EH9 3FD, United Kingdom.}
\author{Ben M. Guy}
\affiliation{School of Physics and Astronomy, The University of Edinburgh, King's Buildings, Peter Guthrie Tait Road, Edinburgh, EH9 3FD, United Kingdom.}
\author{Guilhem Poy}
\affiliation{Universit\'e de Lyon, \'Ecole Normale Sup\'erieure de Lyon, Laboratoire de physique, 46 All\'ee d'Italie, 69364 Lyon, Cedex 07, France.}
\author{Michael E. Cates}
\affiliation{DAMTP, University of Cambridge, Centre for Mathematical Sciences, Wilberforce Road, Cambridge CB3 0WA, United Kingdom.}
\author{Matthieu Wyart}
\affiliation{EPFL SB ITP PCSL, BSP 29 (Cubotron UNIL), Rte de la Sorge, CH-1015 Lausanne,
Switzerland.}
\author{Wilson C. K. Poon}
\affiliation{School of Physics and Astronomy, The University of Edinburgh, King's Buildings, Peter Guthrie Tait Road, Edinburgh, EH9 3FD, United Kingdom.}
\date{\today}

\begin{abstract}
We present experimental results on dense corn-starch suspensions as examples of non-Brownian, nearly-hard particles that undergo continuous and discontinuous shear thickening (CST and DST) at intermediate and high densities respectively. Our results offer strong support for recent theories involving a stress-dependent effective contact friction among particles. We show however that in the DST regime, where theory might lead one to expect steady-state shear bands oriented layerwise along the vorticity axis, the real flow is unsteady. To explain this, we argue that steady-state banding is generically ruled out by the requirement that, for hard non-Brownian particles, the solvent pressure and the normal-normal component of the particle stress must balance separately across the interface between bands. (Otherwise there is an unbalanced migration flux.) 
However, long-lived transient shear bands remain possible.
\end{abstract}

\maketitle
\section{Introduction}

Newtonian liquids, such as water, ethanol or honey, are each characterized by a well defined, shear-rate-independent 
viscosity $\eta$.
In contrast, many complex fluids, such as particulate
suspensions, surfactant solutions and polymer solutions, show shear thinning
and/or shear thickening, so that their steady-state viscosity depends on the
shear rate $\gdot$. The shear
stress plotted as a function of shear rate $\sigma(\gdot) = \eta(\gdot)\gdot$, known as the flow curve, then has a slope less than or greater than unity, for shear thinning or shear thickening respectively, when plotted on logarithmic axes.

In extreme shear thinning or thickening systems, there can in principle appear
regions of the flow curve where $\partial \sigma(\gdot)/ \partial \gdot < 0$ for a range of flow rates. Homogeneous flow is then mechanically unstable \cite{olmsted2008}. 
In many such cases there exist inhomogeneous, shear-banded states that allow the system to flow steadily in time despite this instability. These involve either bands oriented layerwise along the vorticity direction with the same shear rate but different shear stresses (vorticity banding), or bands oriented in the gradient direction
with the same shear stress but different shear rates (gradient banding). There are cases, however, where such banded flows are themselves unstable, giving rise to time-dependent flows with fluctuating shear stresses and rates. These unsteady flows vary from relatively simple oscillations to fully-developed chaos; shear-band-like features may or may not remain detectable. 

These chaotic flows arise from viscoelastic instabilities at essentially zero Reynolds number (negligible inertia), in contrast to conventional fluid turbulence; they are sometimes called `rheochaos'. Viscoelastic instabilities are relatively well studied in entangled micellar systems. In that context, they are often interpreted in terms of an interplay between a slow fluid relaxation time (Maxwell time, $\tau_M$) and an even slower process that modulates $\tau_M$ \cite{cates2006}. However, rheochaos can equally arise in systems without this timescale separation, such as simple nematic fluids \cite{chakrabarti2004}.

In this paper, we present detailed experimental evidence for unsteady flow (leading to rheochaos) in a shear-thickening suspension, and explore the various regimes that emerge. The suspension is granular, rather than colloidal, comprising particles that are large enough for Brownian motion to be negligible. Its flow curve is predicted theoretically to be non-monotonic, in a way that might normally be expected to support steady shear bands. Without Brownian motion, however, we will argue that 
such bands are generically disallowed, so that the flow is unsteady.

Perfectly hard spheres have functioned as a conceptual model for the rheology of particulate suspensions for a long time \cite{einstein1905,batchelor1977}, and continue to yield many insights, e.g., in the study of viscosity divergence near glassy arrest \cite{Gotze}. As two idealized hard spheres approach each other through a fluid with no-slip boundary conditions, the time taken to drain the layer of fluid between them (the lubrication film) diverges, and large `lubrication forces' prevent the particles from ever making contact. 
However, if the particles are slightly rough, or have a finite slip length, they can come into contact when the lubrication film reaches a thickness comparable to the surface roughness, or the slip length. In practice therefore, direct contact forces certainly play a role in real ``hard-sphere" suspensions \cite{gadalamaria1980}, and these contact forces can be expected, in general, to include static friction.

Surprisingly, at low volume fraction, $\phi$, the viscosity of a suspension of  spheres in frictional contact is \emph{lower} than that of an identical suspension of smooth spheres \cite{wilson2000}.
The opposite holds at high $\phi$, where frictional contacts have recently been demonstrated by experiments \cite{guy2015,lin2015}, simulations \cite{Seto2013,Mari2014,Mari2015,ness2016} and theory \cite{Wyart2014} to play a crucial role in shear thickening, and ultimately jamming. 

These recent advances formalize and develop earlier insights from  Melrose and Ball (MB). In their simulations of non-Brownian spheres \cite{melrose1995}, MB found that the gap between the surface of particles, $d$, could fall to molecular dimensions in a real suspension, giving rise to numerically diverging lubrication forces (which scale as $d^{-1}$). MB overcame this problem by introducing a short range repulsive force that
stops these pathologically small gaps from forming \cite{melrose1996}.
In a real suspension, such repulsion can arise from stabilising polymers and/or charges on particle surfaces. Significantly, MB pointed out that the `small gap problem' would recur above a stress $\sigma^\star$ at which the stresses in the system overcome these stabilising repulsions. Crucial to recent advances is the realization that, when the force threshold for a particular contact is exceeded, its lubrication film may fail immediately (due to roughness or a finite slip length) allowing the particles to come into direct frictional contact.
Thus the `onset stress' $\sigma^\star$ marks a crossover from open, well lubricated (or sliding) contacts between particles to direct, frictional (or rolling) contacts.

Developing this insight, Wyart and Cates \cite{Wyart2014} (WC) have constructed a phenomenological theory for the steady flow of 
shear-thickening particulate suspensions. They take all particle interactions to be lubricated
(frictionless) when
$\sigma \ll \sigma^\star$, so that the system is quasi-Newtonian with a viscosity
that diverges at random close packing $\phi_0 \approx 0.64$. However, when $\sigma \gg \sigma^\star$, all contacts are
frictional and the system is again quasi-Newtonian, but now with a viscosity diverging
at some lower volume fraction, $\phi_m < \phi_0$, whose value depends on the
inter-particle static friction coefficient, $\mu_p$ \cite{Mari2014}. The transition
between these two regimes on increasing $\sigma$ causes
shear thickening. 

This scenario resolves a longstanding puzzle in dense suspension theory. Strictly hard particles can have no stress scale $\sigma^\star$, and, without Brownian motion, also have no time scale at rest. Hence, all stresses must scale linearly with $\gdot$ \cite{morris2009}. 
(This includes nonvanishing normal stresses, which is why we describe the two limiting branches as quasi-Newtonian, not Newtonian.) Thus, in the absence of Brownian motion and inertia, 
shear-thickening requires some deviation from strict hard sphere behavior. The key idea of recent work is that this deviation provides, in effect, a stress-dependent inter-particle friction \cite{Seto2013,Fernandez2013,Wyart2014}.

In this paper, we study the rheology of corn-starch suspensions below and above $\phi_m$. Above $\phi_m$ complete jamming is expected, surprisingly, however, we show that flow is still observed, 
but is always unsteady, and shows rheochaos at high enough stress. This relates to the fact that you can run, but not stand still, on a pool of corn starch. Similar unsteadiness is seen for $\phi_c<\phi<\phi_m$, so that the entire DST region is affected. After describing these results we give arguments that unsteady flow should, on theoretical grounds, be a generic feature of dense particulate suspensions in the DST regime.

\section{Methods and set up}
\begin{figure}[!hbt]
\includegraphics[width=0.48\textwidth]{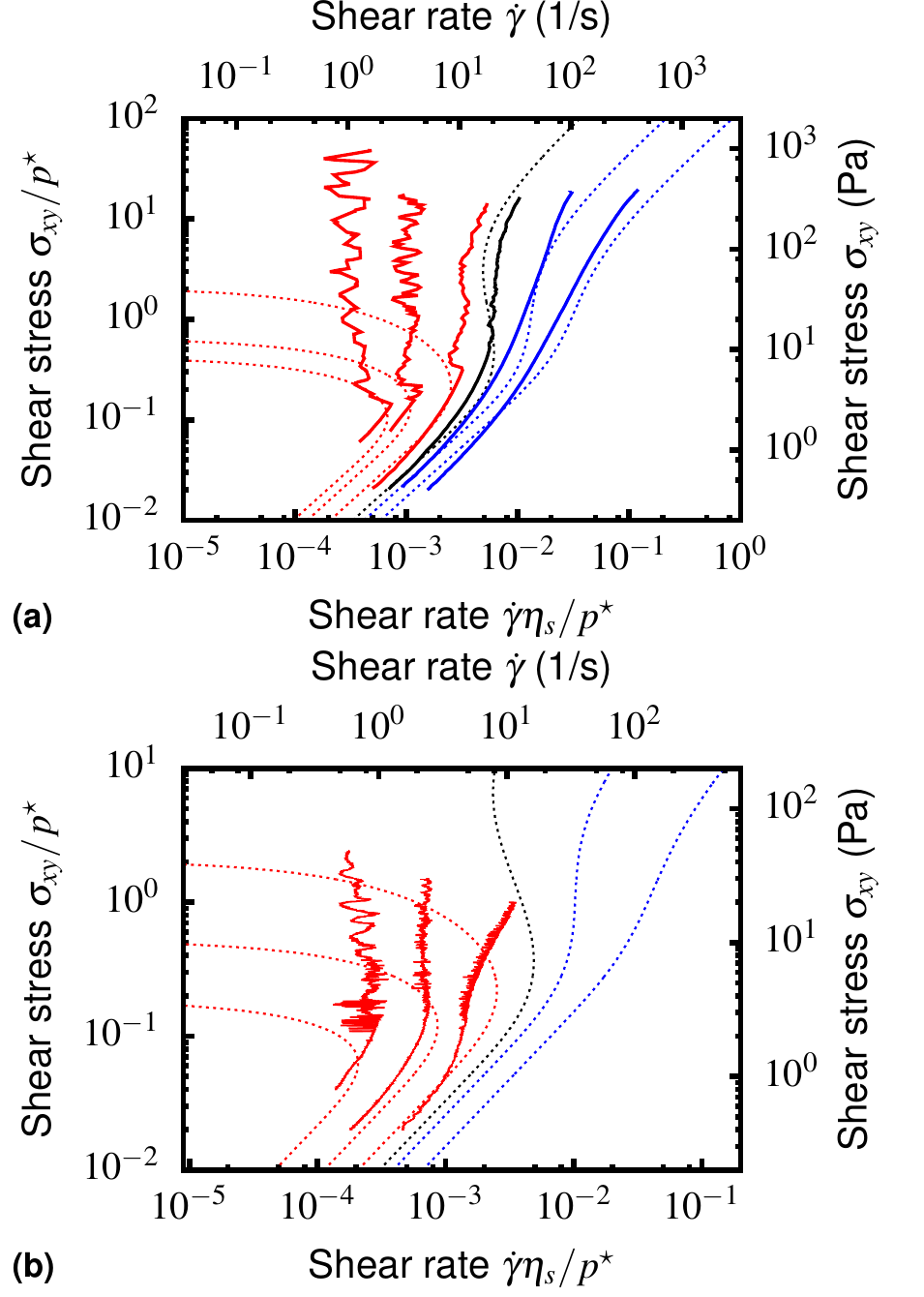}
\caption{(a) Apparent shear stress $\sigma_{xy}$ \emph{vs} rim shear rate $\gdot$ for
corn starch suspensions at mass fractions $\phi^{\rm w}=$ 0.45, 0.46, 0.465, 0.47, 0.50 and 0.52 from right to left. Data represent upward stress sweeps measured between hatched plates. Stress is
reported in Pa (right vertical axis) and in units of the onset pressure for shear
thickening, $p^\star=20.0$ Pa (left vertical axis). Shear rate is reported in
s$^{-1}$ (top horizontal axis) and reduced units $\gdot \eta_s/p^\star$
(bottom horizontal axis). Dashed lines: prediction of \autoref{eq:eta} at
different $\phi$ (0.50,0.525,0.54,0.565,0.585 and 0.595) with $\phi_m=0.55$ and $\phi_{\textrm{RCP}}=0.66$;
these volume fractions were chosen to match experimental data.
(b) The same as above but measured using a Couette geometry, mass fractions $\phi^{\rm w}=$ 0.47, 0.50 and 0.53 from right to left. The dashed lines are predictions from theory for $\phi=$ 0.49, 0.53, 0.545, 0.565, 0.595 and 0.615.
\label{fig:guilhem}
} 
\end{figure}

Rheological measurements were performed with a stress-controlled rheometer
(DHR-2, TA Instruments) with hatched parallel plates, $R=40$ mm diameter
(\autoref{fig:guilhem}(a) and \autoref{fig:jam}) or with a Couette cell (inner
diameter of 18 mm, outer diameter 21 mm) and roughened boundaries
(\autoref{fig:guilhem}(b)) at a temperature of \SI{20}{\degree C}. 
We obtained the raw torque and strain data at a rate of $\sim 10^3$ Hz using
the TA software tool ``ARG2AuxiliarySample". The sample was imaged from the
side using a digital camera at a frame rate of 30 fps using a 16 mm macro
objective.

We performed experiments on corn starch (Sigma Aldrich, unmodified regular corn
starch containing approx.~73\% amylopectin and 27\% amylose [S4126]; diameter
$\approx 14$ $\mu$m, polydispersity $\approx 40\%$ from static light
scattering) dispersed in a mixture of 50wt\% water and 50wt\% glycerol
(viscosity $\eta_s =6$ mPa$\cdot$s, density $\rho_s =1.17$ g$\cdot$cm$^{-3}$)
at various concentrations. The particles swell in our solvent, so that we
cannot access the volume fraction $\phi$. We therefore quote mass fractions
$\phi^{\rm w}$.  Samples were freshly mixed for each experiment and rested for
several minutes before loading into the rheometer. Sedimentation and
evaporation begin to influence the rheology after $\sim 30$ min with parallel
plates; we discard data taken after this time.

Flow curves, \autoref{fig:guilhem}, were obtained by increasing the torque,
$M$, continuously with a logarithmic rate from 0.1 Pa to 1000 Pa over 300 s.
Most samples show edge fracture at stresses between 100 Pa and 1000 Pa, we do not show
any data points for which this has happened. In parallel plate work we report
the shear rate at the rim of the plates $\gdot=R\dot{\varphi}/H$, where $H$ is
the gap height, and the apparent shear stress, $\sigma=2M/(\pi R^3)$. 

\section{Results}

\begin{figure*}[!hbt]
\includegraphics[width=0.7\textwidth]{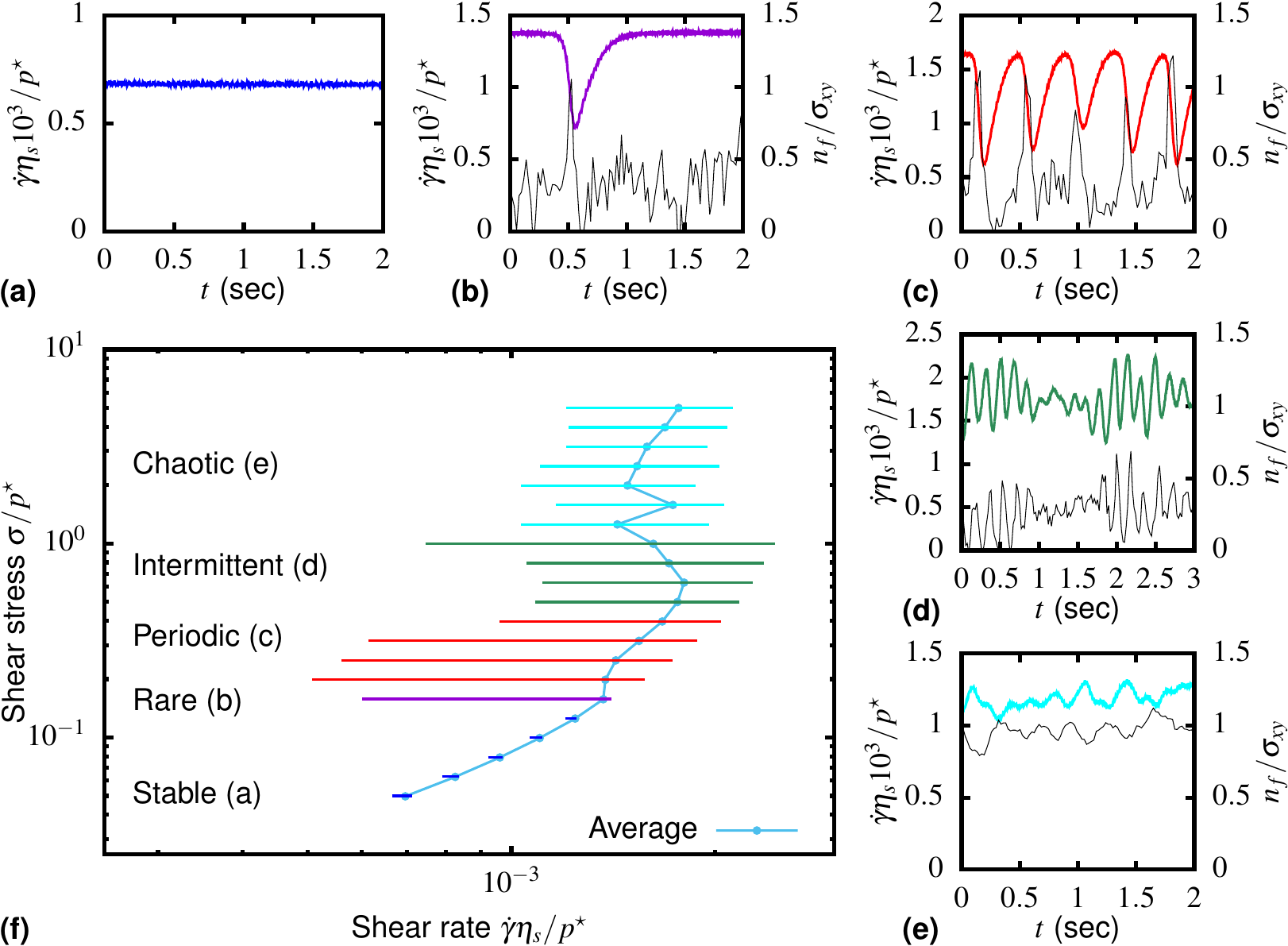}
\caption{ 
(a)-(e) Apparent shear rate as a function of time for increasing stress, on the left y axis. 
The thin black lines show the normal pressure $n_f/\sigma_{xy}$ on the right y axis.
(f) Apparent shear stress as a function of rim shear rate $\gdot_R$ in
absolute- and reduced-units for corn starch at a mass fraction of $\phi^{\rm w}=0.52$,
corresponding to a volume fraction just above $\phi_m$ in WC theory. Horizontal
lines: raw $\gdot_R$ data at different applied $\sigma_{xy}$ in the stable
(dark blue), periodic (red), intermittent (green) and chaotic (cyan) regimes.
Symbols: average $\gdot_R$. 
\label{fig:jam}
} \end{figure*}

\autoref{fig:guilhem} shows flow curves measured at different mass fractions,
$\phi^{\rm w}$ (see caption), reported as the reduced shear stress
$\sigma/p^\star$, versus the reduced shear rate, $\gdot \eta_s/p^\star$. Here
$p^\star$ is the onset pressure for the formation of frictional contacts, related to the 
onset stress through $\sigma^\star=\mu(\phi)p^\star$, see Section \ref{sec:theory}.
(Note that we control the shear stress, plotted on the vertical axis, and measure the shear rate, on the horizontal axis.)
At $\phi^{\rm w}<\phi^{\rm w}_c\approx 0.465$, we observe continuous shear
thickening above an onset pressure $p^\star = 20.0\pm5$ Pa to a
high-viscosity quasi-Newtonian state (blue curves in \autoref{fig:guilhem}).
The steepness of the shear-thickening part of the flow curve increases with
$\phi^{\rm w}$ until, at $\phi^{\rm w}_c$, $d\gdot/d\sigma=0$ beyond which the
sample discontinuously shear thickens. In contrast to the continuous case,
where the flow is steady throughout the flow curve, we now observe large
shear-rate fluctuations above the critical stress, resulting in considerable
spread in the data. These fluctuations are also present in constant stress
experiments (as shown in \autoref{fig:jam}) and remain present for long times
(at least 30 min). Such large fluctuations arise as soon as the measured flow
curve starts bending backwards. 
(Hence there is no inconsistency in the apparent negative slopes of the empirical, averaged, `flow curves'.)

\begin{figure}[!hbt]
\includegraphics[width=0.30\textwidth]{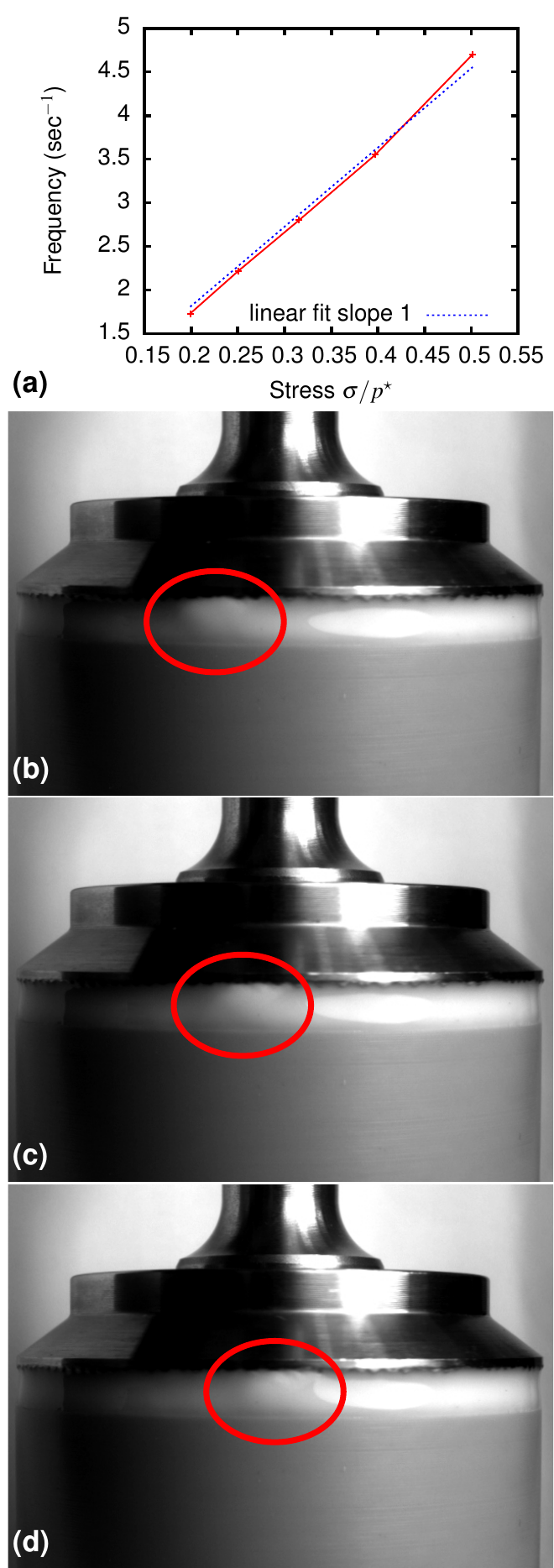}
\caption{ 
(a) The frequency of the oscillation as a function of the applied stress for a sample at $\phi^W=0.50$ measured between hatched plates.
(b)-(d) Pictures of the deformation of the interface during the intermittent regime d at $\phi^W=0.50$ between hatched plates. 
In the photos the front of the cone rotates to the left, the gray bottom plate is stationary, and
a deformation of the interface (highlighted by a red ellipse) moves to the right. 
\label{fig:photo}
} \end{figure}

Just above $\phi^{\rm w}_c$ (black curve, \autoref{fig:guilhem}(a), measured
between hatched parallel plates), there is a narrow concentration range in
which the system can reach a flowing quasi-Newtonian state at high stresses, as
previously reported \cite{Bender1996}, although we observe severe
deformations of the meniscus in this regime. Above a second critical
concentration $\phi^{\rm w}_m \approx 0.47$ (red curve,
\autoref{fig:guilhem}(a) and (b)), no such quasi-Newtonian regime is found even
at the highest observable stresses; instead the flow is always erratic. 
We observe very similar behaviour in a Couette geometry, \autoref{fig:guilhem}(b). 
These time-averaged observations map rather directly onto the WC theory of steady-state shear thickening
if we identify $\phi^{\rm w}_c$ with  $\phi_c$, the point where sigmoidal flow
curves emerge, and $\phi^{\rm w}_m$ with $\phi_m$, the jamming point for
frictional particles. On the other hand, the theory does not capture the
magnitude of the shear thickening completely, most likely due to the wide size
and shape dispersity in corn-starch, or non-hard interactions, which also 
give rise to a small yield stress (not shown).

Significant differences between experiments and theoretical expectations (see section \ref{sec:theory}) arise for $\phi > \phi_m$. Here, WC theory leads us to expect that no steady flow is possible above a threshold of stress, even with shear bands present, because there is no upper branch to the flow curve. However, at low stresses, steady flow is possible on the lower branch, but beyond it the only steady state either has coexistence of low and high stress bands, both at $\gdot = 0$, or is jammed homogeneously (again with $\gdot = 0$). Thus one might expect the system to be able to support a relatively modest static load without flowing at all.

\begin{figure*}[!hbt]
\includegraphics[width=0.95\textwidth]{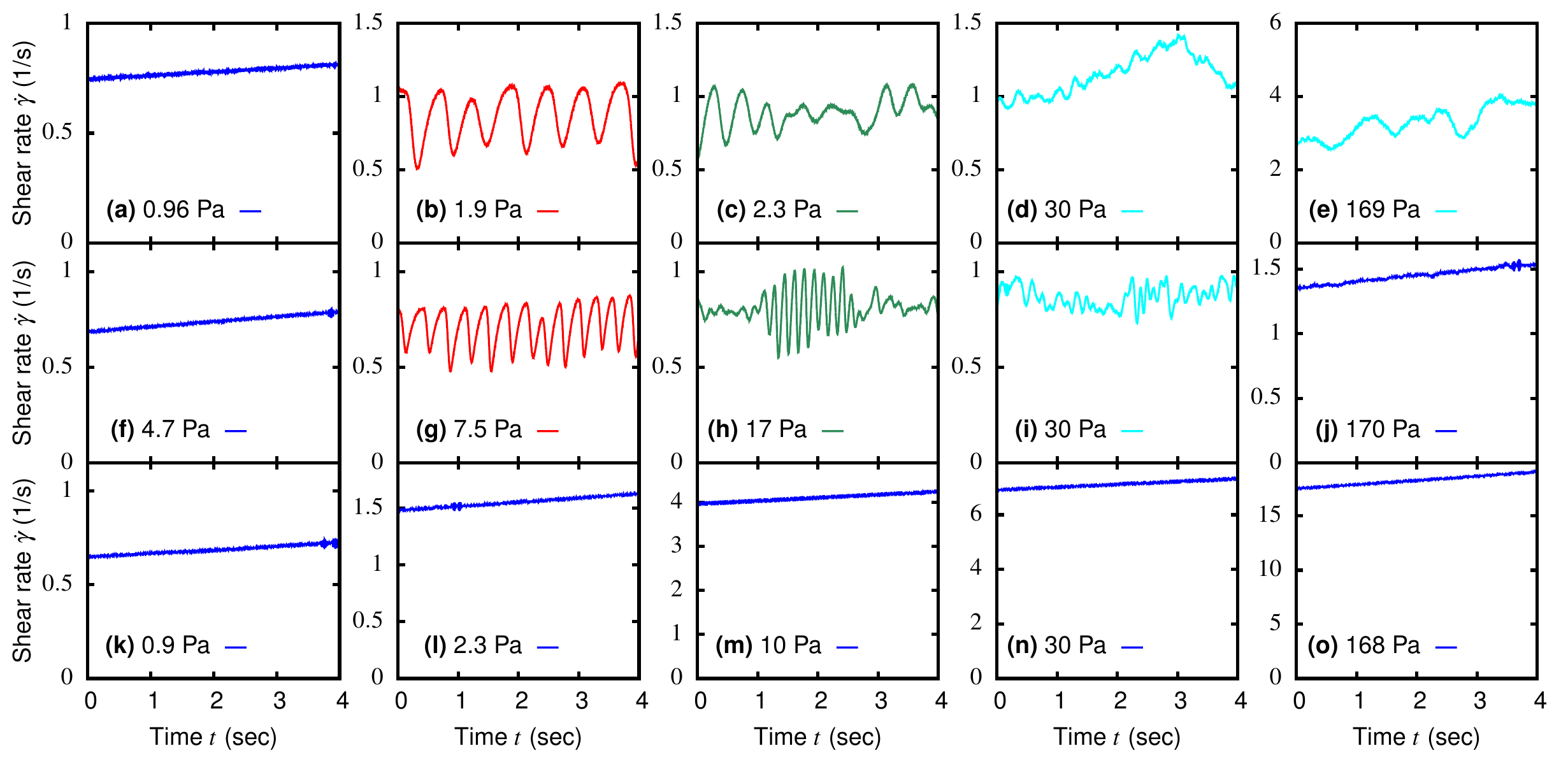}
\caption{ 
(a)-(e) Apparent shear rate as a function of time measured during a continuous increase in shear stress in a Couette geometry at $\phi^W=0.50$.
(f)-(j) Same as a-e but in a parallel plate geometry at $\phi^W=0.465$.
(k)-(o) Same as a-e but in a parallel plate geometry at $\phi^W=0.45$.
\label{fig:gamma}
} \end{figure*}

However these WC scenarios refer to steady states. Experimentally we find instead that the system \emph{does} flow at high stresses in this regime, but flows unsteadily. 
The phenomenology of this `unexpected' flow at $\phi > \phi_m$ is complex. To begin to explore it, \autoref{fig:jam}(f) shows the time-averaged flow curve, as well as the measured fluctuations, in a sample at $\phi^{\rm w}=0.50$, corresponding to a volume fraction just above $\phi_m$. 
At the lowest applied shear stresses, $\sigma <0.1 p^\star$, 
the shear rate fluctuates only a little around a well-defined average (see \autoref{fig:jam}(a)). The axial stress measured on the top plate, $N$, is close
to the noise level of the transducer\footnote{Note that the stress measured on the top plate is a combination of the of the first and second normal stress differences and not a measurement of the particles 
pressure as measured by \cite{Boyer2011}.}.
The meniscus at the air-sample interface remains smooth, shiny and undisturbed. 
We observe a drift in the shear rate after long times (hours), presumably due to particle migration, sedimentation or evaporation. 

For $0.1\sigma^\ast \lesssim \sigma \lesssim 0.2p^\star$, 
region B in \autoref{fig:jam}(f), the flow is steady for seconds, but is punctuated by sudden drops in $\gdot(t)$, \autoref{fig:jam}(b). 
We refer to these events as ``jams", and argue that they are related to the formation of locally-solid regions within the suspension. 
During a jamming event, $\gdot$ (purple and red lines) drops rapidly, with a concomitant positive spike in the axial stress (black lines), before increasing slowly back to the steady-state value. 

While the jamming events in region B are sparsely distributed and seem to occur randomly in time, they become very regular with a well-defined frequency at $\sigma \gtrsim 0.2 p^\star$, 
regime C \autoref{fig:jam}(f). 
This is visible macroscopically as periodic jerks of the rheometer top plate. 
The minimum shear rate reached during a jamming event is variable, \autoref{fig:jam}(c), while the shear rate in the flowing state is approximately the same and corresponds to the right-hand limit of the horizontal lines in \autoref{fig:jam}(f). 
These oscillations remain over long times and only change over the course of hours (presumably as the sample dries out).
The frequency of the oscillations increases linearly with the applied stress, \autoref{fig:photo}(a).
Each sudden decrease in $\gdot$ is accompanied by a localised deformation of the air-sample interface.
A small area of the interface comparable to the gap height bulges out slightly, while the surrounding area curves slightly inward. The interface recovers a smooth profile as the plate accelerates back to the steady state value. 
Note that these localised jams are not an artifact of the cross-hatched plates; they start to appear at the same stresses with smoother surfaces, albeit in the presence of significant wall slip, as well as in Couette geometries (\autoref{fig:gamma}(b)). 

In region D, \autoref{fig:jam}(f), periodic jamming coexists temporally with bursts of unpredictable fluctuations, as shown in \autoref{fig:jam}(e). During the periodic intervals, the air-sample interface behaves the same as in region C, with short-lived, static jammed regions appearing at the same time as the drop in shear rate. During the random bursts, more irregular surface deformations are observed that are long-lived and move around the interface opposite to the direction of flow (see \autoref{fig:photo}b-d). 
Usually, only one or two transient deformations appear during each intermittent event and disappear when the periodic oscillations resume.

At the highest stresses $\sigma/p^\star \gtrsim 1$, 
in region E, \autoref{fig:jam}(f), the periodic jamming and unjamming is absent, and only random-looking fluctuations are observed, \autoref{fig:jam}(e). This behavior, and the series of events at lower stresses that precede it, are similar to the development of rheochaos as observed in micellar systems \cite{cates2006}.
We leave it to future work to establish whether the flow is really chaotic in a technical sense; for our purposes what matters is that it is unsteady, not readily predictable, and without obvious periodic features.
In region E, the first normal stress difference is permanently large and positive and anti-correlated with the shear rate. 
Very recently, unstable flow, sudden jams and a transition to what appears to be rheochaos have been observed in 2D computer simulations of inertial frictional grains \cite{grob2015}. Although the origin of the sigmoidal flow curves is different, the types of unstable flow observed there are very similar to the ones reported here. 

We observe the same transition sequence in a Couette geometry as with parallel plates, although the onset stress for unsteady flow is lower in a Couette geometry than between parallel plates \autoref{fig:gamma}(a-e). 
We observe the same sequence of phenomena for other volume fractions above $\phi_m$, whereas for samples just below $\phi_m$ we observe an additional steady flow regime at high stress, \autoref{fig:gamma}(j). 
At lower volume fractions we do not observe shear rate fluctuations at any applied stress \autoref{fig:gamma}(k-o).

\section{Theory} \label{sec:theory}

\begin{figure*}[hbt!]
\includegraphics[width=0.98\textwidth]{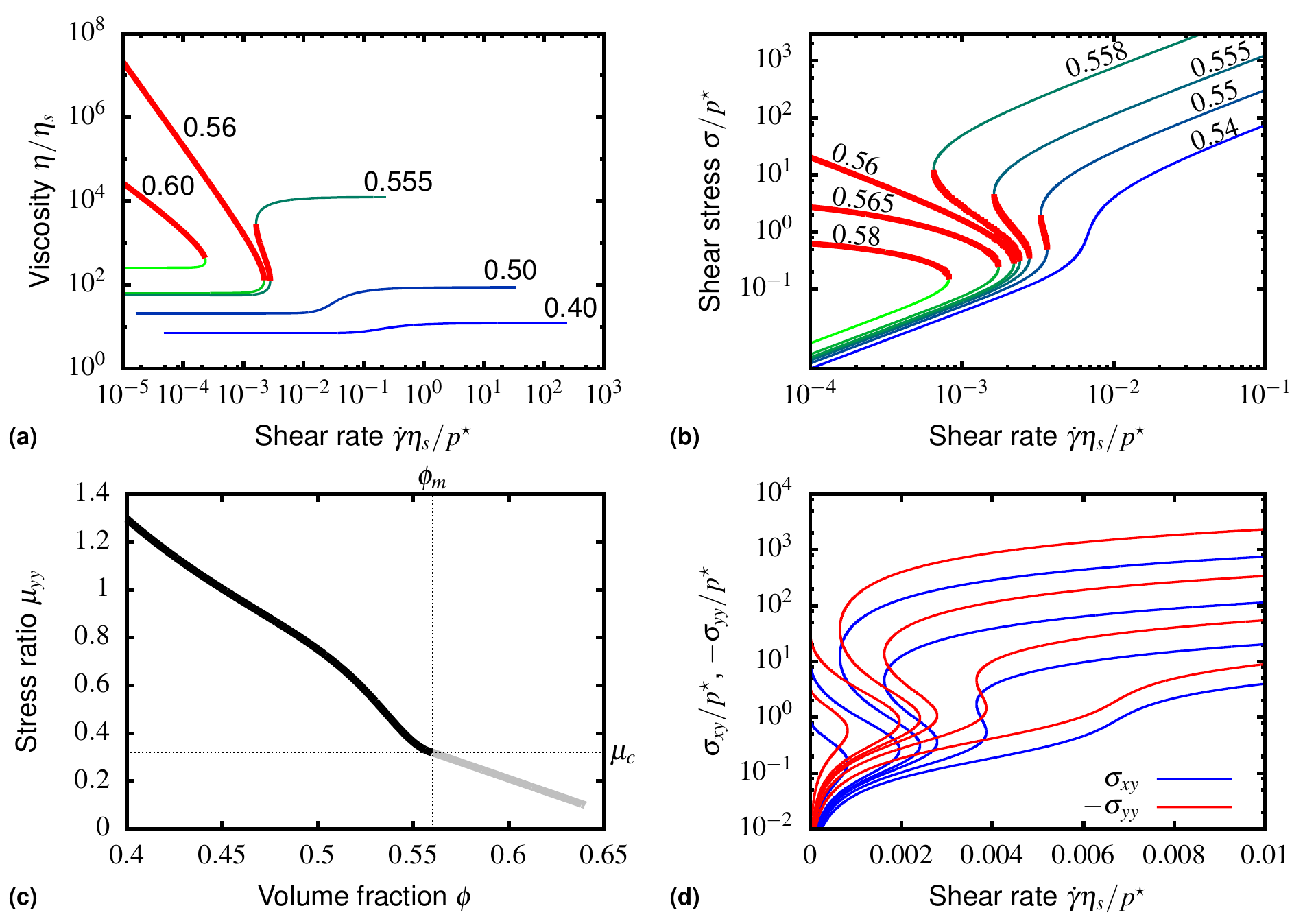}
\caption{
(a) Relative viscosity $\eta/\eta_s$ \emph{vs} reduced shear rate $\gdot \eta_s/p^\star$ at
different volume fractions $\phi$ (as labelled) predicted by the theory of Wyart and Cates
\cite{Wyart2014}, (\autoref{eq:eta}). We take $\phi_m=0.56$, $\phi_\mathrm{RCP}=0.64$,
$\beta=1$ from recent experiments \cite{guy2015}. 
The unstable regimes are marked in thick (red) lines.
(b) Corresponding flow curves (shear stress as a function of shear rate).  
(c) The $\phi$-dependent stress ratio (or effective macroscopic friction coefficient) in the $y$- (gradient-)
direction, $\mu_{yy}=\sigma_{xy}/\sigma_{yy}$, used to obtain these plots. Black solid line: derived from Boyer \emph{et al.} \cite{Boyer2011}, applicable up to $\phi_m$; grey line: a plausible
extrapolation to higher $\phi$ based on 2-d simulations \cite{trulsson2012}. (d) The flow curves  $\sigma_{xy}(\gdot)/p^\star$ at different $\phi$ (blue
lines) plotted again, now against a linear horizontal axis, and compared with the normal stress in the $y$-direction $-\sigma_{yy}(\gdot)/p^\star$ (red lines) calculated using the expression for $\mu$ shown in (c).
\label{fig:fc}} 
\end{figure*}

In this section we will summarise the steady-flow theory outlined by \cite{Wyart2014}. We will then explore what this means for the stability of the flow of shear thickening suspensions. 
WC describe the rheology of dense non-Brownian suspensions with a jamming
volume fraction, $\phi_J(p)$, that depends on $p$, the particle pressure, defined via the trace of the particle contribution to the stress. This $\phi_J(p)$ evolves smoothly from $\phi_J(0) = \phi_0$ to
$\phi_J(\infty) = \phi_m$ as the fraction of frictional contacts $f$ goes from 0 to 1:
\begin{equation}
\phi_J(p)=\phi_m f(p/p^\star)+\phi_\mathrm{0} [1-f(p/p^\star)].
\label{eq:phiJ}
\end{equation}
Here $f$, which is dimensionless, can depend only on the ratio of $p$ to the onset stress, as written above. The precise form of $f$ is inessential, but a stretched exponential
\begin{equation}
\label{eq:exp}
f=\exp{\left[(-p^\star/p)^{\beta}\right]},
\end{equation}
gives good agreement with experiments \cite{guy2015} and simulations. 

At the macroscopic level, the particle pressure $p$ is related to the shear stress $\sigma$
through a stress ratio or macroscopic friction coefficient $\mu(\phi)$ (not to be confused with $\mu_p$ as defined above): 
\begin{equation}
\sigma = \mu(\phi) p,
\end{equation}
where $\mu$ is taken by WC to depend only on $\phi$.
This involves a simplification, since in principle the macroscopic friction coefficient $\mu$ could certainly also depend on the state of microscopic friction and hence on $f$ \cite{lemaitre2009,Boyer2011,Wyart2014}.
We return to this issue below.
This relation between stress and pressure allows us to write \autoref{eq:phiJ} as function of stress instead of pressure 
\begin{equation}
\phi_J(\sigma)=\phi_m f(\sigma/\sigma^\star)+\phi_\mathrm{0} [1-f(\sigma/\sigma^\star)],
\end{equation} where $\sigma^\star=\mu(\phi)p^\star$.

Finally, the suspension viscosity $\eta = \sigma(\gdot)/\gdot$ is known to diverge as the jamming transition is approached \cite{Boyer2011}.
This divergence is related to the explosion of velocity fluctuations caused by excessive crowding \cite{lerner2012,andreotti2012}, and can be computed in simple models \cite{degiuli2015}.
In WC this effect leads to a divergence of viscosity at $\phi_J(P)$ modeled as
\begin{equation}
\eta(\sigma,\phi)=\sigma/\gdot=\eta_s\left[1-\frac{\phi}{\phi_J(\sigma/\mu(\phi))}\right]^{-\alpha},
\label{eq:eta}
\end{equation}
with an exponent estimated as $\alpha=2$. This leads to $S$-shaped flow curves, whose likely role in shear thickening was earlier identified by \cite{bashkirtseva2009}.

\autoref{fig:fc}(a) shows the reduced suspension viscosity, $\eta(\sigma,\phi)/\eta_s$, predicted by the WC model as a function of reduced
shear rate $\gdot \eta_s/\sigma^\star$ using $\phi_0=0.64$,
$\phi_m=0.56$ and $\beta=1$. For $\phi$ somewhat less than $\phi_m$,
the system shear thickens continuously between the two quasi-Newtonian regimes. The slope
$d\eta/d\gdot$ increases with $\phi$ until, at a critical $\phi=\phi_c \approx
0.55$, $\eta(\gdot)$ becomes vertical. For $\phi>\phi_c$,
$\eta(\gdot)$
contains a region of negative slope and develops a sigmoidal shape, while tending
towards quasi-Newtonian regimes at both low and high stresses. Above a second critical volume
fraction set by $\phi=\phi_m$, the backward-bending part of the flow curve meets the vertical axis and there is no longer a flowing branch at high stresses. The corresponding $\sigma (\gdot)$ curves are shown in \autoref{fig:fc}(b).

As $\phi$ is increased, the theory predicts first continuous shear thickening, then discontinuous shear thickening (DST) between two flowing branches each of finite viscosity \cite{Bender1996}, and finally DST from a flowing branch to a jammed branch that cannot flow at finite $\gdot$ without some sort of fracture \cite{Laun1994}. 
This last regime, which arises for $\phi > \phi_m$, is called `complete jamming' \cite{cates2005};
in it, the putative upper branch of the flow curve $\sigma(\gdot)$ runs straight up the vertical axis. 
The WC model fits recent $\eta(\sigma,\phi)$ data on suspensions of sterically stabilized polymethylmethacrylate (PMMA) particles whose interactions closely approach the hard-sphere limit \cite{guy2015}.  The predicted sigmoidal flow curves, although pre-empted by instability in bulk steady flows, have since been observed, at least transiently, in experiments and simulations of nearly-hard non-Brownian particles \cite{pan2015,Mari2015}.  

At high volume fractions, in the complete jamming regime, the WC theory requires that any high-stress shear band must have zero flow rate, $\gdot=0$. This is because the only other frictional states on the flow curve have $d\sigma/d\gdot<0$ and are themselves unstable. Thus, any steady banded state comprises coexistence, layerwise along the vorticity direction, of a jammed state at finite stress and a fluid state at zero stress (since with this orientation, $\gdot$ is equal in both bands). Thus, no steady flow is possible even with shear bands present; the only steady flow states for $\phi>\phi_m$ are homogeneous and lie on the low-friction branch. Dynamically, if the mean shear stress is increased beyond the stability limit of that branch, one might then expect its local value to become increasingly heterogeneous along the vorticity direction until flow stops altogether for the reasons just described.

\section{Absence of Steady Shear Bands}

It is notable that in our experiments, we observe unsteady flow at all
concentrations $\phi > \phi_c$ where stable banded flow may, at least at first
sight, be expected. We now turn to explore the origins of these instabilities
in our system. 

Flow instability, oscillation, and rheological chaos has been fairly widely
reported in both shear-thinning and shear-thickening viscoelastic materials
(particularly but not exclusively micellar solutions \cite{cates2006}).  Given
the presence of highly nonlinear constitutive equations  that relate stress to
strain-rate history, one might expect instability to be more common.
Mathematically, unsteady solutions can either arise `directly' from the
instability of a steady homogeneous flow, or through a similar instability
within one of the shear-bands that would otherwise allow steady but
inhomogeneous flow \cite{fielding2004}.

Although in general one does not expect simple rules to govern whether flows
are steady or unsteady, dense non-Brownian shear-thickening suspensions present
a somewhat special case in relation both to vorticity bands and to gradient
bands.
Below we deal with these two cases in turn. We consider the case
where the flow curve has a sigmoidal shape, which occurs for $\phi_c \leq \phi
< \phi_m$, as well as the regime $\phi > \phi_m$ (with no upper flow branch), which applies in most of the experiments presented above. 
We refer to the flow-, gradient- and vorticity-directions as $x$, $y$ and $z$, respectively. 

Let us consider the diagonal components of the stress tensor, which comprise an isotropic solvent pressure $-p_s\delta_{ij}$ plus the three normal stresses $\sigma_{uu} = - p_{uu}$ caused by the presence of particles. 
(Here $u = x,y,z$ is a generic, but not summed, Cartesian index; recall the stress and pressure tensors have opposite signs.)
For strictly hard spheres with fixed frictional properties, each normal stress is linear in $\gdot$. More generally we are dealing with a manifold of steady states in which the ratio of shear to normal stresses (i.e., the macroscopic friction constant) depends on both volume fraction and the state of contact friction captured by $f(\sigma)$:
\begin{equation}
-\sigma_{uu} = p_{uu} =\sigma_{xy}/\mu_{uu}(\phi,f(\sigma)).
\label{eq:mu}
\end{equation}
Generically, the $\mu_{uu}$ are unequal, causing normal stress
differences $N_1 = \sigma_{xx}-\sigma_{yy}$, and $N_2=\sigma_{yy}-\sigma_{zz}$.

A simplification made by WC was to suppress the dependence of $\mu_{uu}$ on the
state of contact friction, so that it is a function of $\phi$ only. This is
pursued in \autoref{fig:fc}(c),(d) where $\mu_{uu}(\phi)$ is estimated as
described in Appendix \ref{appendixa}, which also gives further information
about what is known of the $\mu_{uu}$'s in granular systems.  More generally,
however, the $\mu_{uu}$ must depend on $\sigma$ via $f(\sigma)$ as well as on
$\phi$; hence the macroscopic friction will have different values on the lower
and upper limiting branches of the flow curve. Therefore each of the normal
stresses has a shear rate dependence $p_{uu}(\gdot)$ that \emph{qualitatively}
resembles the shear stress $\sigma_{xy}(\gdot)$, but is not
\emph{quantitatively} proportional to it as was assumed in \autoref{fig:fc}(d).
This will prove important in the discussion of gradient bands below. First,
however, we address vorticity bands.

\subsection*{Vorticity bands} 

\begin{figure}[hbt!]
\includegraphics[width=0.4\textwidth]{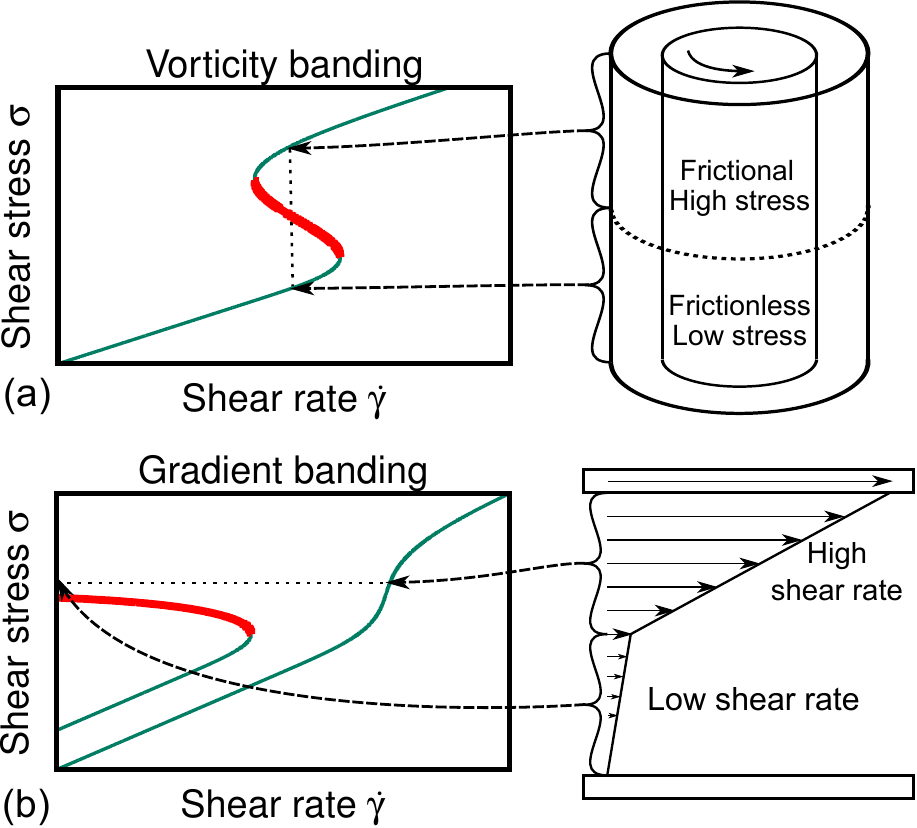}
\caption{
(a) A Schematic of vorticity banding as it could hypothetically occur for a homogeneous volume fraction. 
(b) A schematic of gradient banding as it could hypothetically occur in an inhomogeneous sample. 
\label{fig:schematic}} 
\end{figure}

We consider flow between infinite parallel plates so that homogeneous flow is possible in principle. Steady vorticity bands are expected to arise when the applied steady shear stress $\sigma_{xy}^a$ falls within a window $\sigma_{xy}^{(1)}<\sigma_{xy}^a<\sigma_{xy}^{(2)}$ that includes all part of the flow curve with negative slope. (Vorticity bands are not expected to arise in experiments at controlled $\gdot$ \cite{olmsted2008}.) The vorticity bands have a common shear rate, $\gdot_1=\gdot_2$, but different shear stresses $\sigma_{xy}^{(1)}$ and $\sigma_{xy}^{(2)}$. 
As $\sigma_{xy}^a$ is varied, the fraction of the sample occupied by each type of band adjusts so that the space-averaged shear stress is $\sigma_{xy}^a$. 

Mechanical stability then requires equality between bands of the normal stress
normal to the band interface, $\sigma_{zz}$. We thus have
$p_s^{(1)}+p_{zz}^{(1)} = p_s^{(2)}+p_{zz}^{(2)}$. 
The particle contribution $p_{zz}$ is mediated by forces (perhaps
including lubrication forces) which are in effect transferred directly from
particle to particle through a network of contacts. The fluid pressure $p_s$ is
carried by solvent molecules that can move freely through the pores of
this network. Without Brownian motion to create an osmotic reaction force, any
difference in fluid pressure between bands should drive the solvent to flow
from high to low $p_s$, with mass balance maintained by a flux of particles in
the opposite direction, from high to low $p_{zz}$. 

Thus, $p_s$ and $p_{zz}$ must be separately equal in shear-banded non-Brownian
suspensions. Though the argument is general, it is particularly transparent for
$\phi > \phi_m$, when coexisting vorticity bands are in fact at rest, as
previously explained. No lubrication (or other hydrodynamic) forces then
remain, so the fluid and particle mechanics are completely decoupled. It 
is quite clear in this case that the solvent and particle pressures \emph{must}
be separately equal between bands. The same argument extends to flowing bands,
at least if the system is treated as two continua (solvent and particles) with
a drag term coupling their two velocities (i.e., a two-fluid model)
\cite{andreotti2013}.

Steady vorticity bands thus require not only equal strain rate but also equal
particle pressure $p_{zz}$. Since $f$ is only a function of the particles
pressure $p_{zz}$ (\autoref{eq:exp}) the fraction of frictional contacts and
thus the frictional states of the two bands must be identical (note that $\phi$
itself can be be seen as a function of $f$ and $p_{zz}$ \cite{Boyer2011}).
However, if the frictional state has to be the same in both bands then the
suspension is identical to one with a fixed microscopic friction coefficient
$\mu_p$.  Therefore for vorticity bands to be stable they also need to be
stable for a system with a fixed friction coefficient.  However, suspensions at
fixed friction are well studied and shear banding has not been observed
\cite{lemaitre2009, Boyer2011}.  A supplementary argument, starting from the
same premise and leading to the same conclusion that vorticity banding is
prohibited in dense non-Brownian suspensions, is provided in Appendix
\ref{appendixb}.

It is helpful to discuss separately the case when $\phi>\phi_m$ so that the
bands are not flowing.  The particle stresses in the fluid band vanish. In this
case, equality of $p_{zz}$ would require $\mu_{zz}$ to diverge on the
frictional branch so that there is a large shear stress at vanishing normal
stress.  However, the frictional branch at $\phi > \phi_m$ is a jammed solid.
Such materials can support only a finite stress anisotropy without flowing, so
that $\mu_{zz}$ cannot become infinite as required. Hence coexistence of
non-flowing vorticity bands is ruled out.

\subsection*{Gradient bands} 

We now argue that static \emph{gradient} bands are also ruled out in steady state once particle migration is allowed for. We do not rule these out entirely, but analogous with the vorticity bands, we show that they should only arise under conditions where a system of fixed microscopic friction coefficient would also show gradient banding. 
This statement again relies on the fact that the state of friction, represented by $f$  
can be viewed as a function of $p_{yy}$ only. 
But separate equality of the fluid pressure and the normal-normal stress requires that $p_{yy}$ and hence $f$ is the same in coexisting gradient bands \footnote{Recall that the normal-normal component of the particle stress is (minus) the diagonal component of the stress tensor. Thus if the $z$ direction is the direction normal to the interface the normal-normal component is the $z$-$z$ component of the stress tensor.}.
Accordingly, such bands can only exist if they would also do so in a system of fixed friction. As far as is known, this does not happen for hard particles (but might do, very close to jamming, for deformable ones). A supplementary argument for the prohibition of gradient bands is provided in Appendix \ref{appendixc}.

\subsection*{Discussion}

We have argued that neither vorticity nor gradient banding is generically
sustainable in steady state for dense, shear-thickening suspensions of hard
particles in which mechanical contact and viscous stresses remain unopposed by
Brownian motion.  Avoidance of mechanically-induced particle migration then
requires that the particle normal stress contributions $p_{uu}$, with $u$
normal to the interface between bands, and the solvent pressure $p_s$, are
separately equal in coexisting shear bands. 

This condition holds only in strict steady state where all fluxes between bands
must vanish. Quasi-steady shear bands could however be sustained under
transient conditions by a nonzero flux of particles across the interface. One
possible explanation of the flow-NMR data in \cite{Fall2015}, which apparently
show static gradient-bands, is that these represent a snapshot of the system
while such fluxes remain transiently present \cite{olmsted2008,adams2011}. 
(Note, however, that \cite{Fall2015} used a wide-gap Couette
system. In this geometry banding is expected even for fixed-friction materials
because the imposed ratio of shear to normal stress varies with radius, and may
thus be unrelated to shear thickening. 

The experimental fact, in any case, is that steady flow is not seen in our
system whenever shear-banding would be needed to create it. We have made
similar observations on other materials than corn-starch and we believe this to
be the generic outcome for shear-thickening materials under conditions of
imposed stress. We leave open the question of what to expect under conditions
of imposed strain rate; since in fact only the average strain rate gets
imposed, it is quite possible that an unsteady stress response will again arise
close to the DST transition.

When steady banding is not possible, our experiments suggest that the dynamical
outcome is as follows.  The system jams locally (near the edge of our geometry,
because the stress is largest at the edge of our parallel plates).  The
particles migrate away from the jammed region due to the unequal particle
pressure in the jammed region. It is this local increase in particle pressure
that drives particle migration that also deforms the meniscus. This migration
continues until the pressures balance and locally the flow is no longer
unstable and the system is unjammed.  These jams always form at the edge of our
sample, in a parallel plate geometry, due to the stress gradient over the
sample. This explains how the system is able to flow deep into the regime where
it would be expected to jam. 

While we have ruled out stable bands in suspensions of non-Brownian hard particles, in Brownian suspensions stable shear bands might be possible. 
For stable bands the solvent pressure difference across the interface between the bands needs to be maintained. 
Without this the particle pressures must be equal and the argument for non-Brownian systems forbids stable bands.
In Brownian systems, such as micellar solutions and small hard-sphere colloids, osmotic forces can maintain an osmotic and thus a solvent pressure difference across an interface. 
For micelles it is known that these these unequal pressures cause differences in Laplace curvature at the external menisci of the two bands (which may fail if inequalities become large) \cite{skorski2011}. 
In the non-Brownian suspensions studied here the meniscus deforms (\autoref{fig:photo}), indicating differences in pressure, however, these deformations are not stable. 
To stabilize the frictional shear band the system needs to maintain a higher particle pressure (and thus a lower solvent pressure) in the frictional band than in the frictionless band. 
The mechanism required to stabilise the bands has to push particles from the frictionless band towards the frictional band and it has to do this against the particle pressure.
Although Brownian motion is required it might not be sufficient for the formation of stable bands. 
Equilibrium effects such as diffusion will never push hard colloidal particles against a pressure gradient. 
One way this can happen is through an out of equilibrium mechanism such as flow concentration 
coupling as in shear thickening micelles solutions \cite{helfand1989,cates2006}.
Whether this is also possible for Brownian hard spheres remains to be investigated. 

\section{Conclusion}

The phenomenology of continuous shear thickening (CST) of non-Brownian
suspensions is well described by the WC theory \cite{Wyart2014,guy2015}. In
this work, we have shown that the same applies in the discontinuous
shear-thickening (DST) regime, so long as one allows that the instability
connected with a sigmoidal flow curve need not lead to the formation of
steady-state vorticity bands.  The steady banding picture would give two
regimes with DST; one in which the bands comprise two different flowing states
(the upper and lower quasi-Newtonian branches of the flow curve), at $\phi <
\phi_m$; and one in which both the low-friction and the high-friction branch
are not flowing, at $\phi > \phi_m$.

The latter is a strong prediction of any steady banding hypothesis since it
implies that, above a relatively modest stress threshold $\sim 5\sigma^\star$, a
static load can be supported indefinitely even though only part of the
structure (the jammed bands) are contributing to its support. If true, this
should presumably also be the case in other geometries of inhomogeneous stress,
such as a person standing on a pool of corn starch suspension. If particle
migration did not matter, the person should be able to stand still indefinitely
without sinking in.

Contrary to the expectations based on any hypothesis of time-independent shear
bands, we find that flow (although not steady) is possible even in this regime
of very high density.  The reason that the system is still capable of flow for
$\phi>\phi_m$ is that it only spends part of its time in a jammed state.
Whenever bands are present, particle migration allows the jammed regions to
dilate and unjam. A new jam then forms somewhere else; bands are unsteady, and
a finite average rate of flow is achieved. Even if the local stress exceeds the
highest threshold calculated by WC, beyond which one expects homogeneous
complete jamming rather than shear bands, particle migration into regions of
lower stress will always allow motion to occur.  This concurs with the
observation that a pool of corn starch cannot in fact support a localized
static weight for very long times \cite{kann2011}.  According to our arguments,
however, if this high threshold is exceeded across the entire sample, flow
would finally cease. Hence, although one cannot stand still on an infinite pool
of corn-starch suspension, it should be possible to do so on a finite bucket of
the material.

If  steady shear bands are indeed ruled out by our arguments, the ubiquitous
unsteady flows the we observe stem naturally from the large, unstable,
negative-slope region in the flow curves, predicted by the WC theory at $\phi$
values close to and beyond $\phi_m$. We observed a transition from periodically
jammed via intermittent to rheochaotic flows upon increasing the stress.
Comparable behavior, while differing from system to system in the particular
route to chaos
(e.g., \cite{Ganapathy2006}), is well known for viscoelastic micellar
solutions. It has even been observed for shear-thickening suspensions before,
but was attributed then to wall slip \cite{larsen2014}. This explanation is
ruled out by the fact that we see the same phenomenology with and without
hatched plates. 

Combining the theoretical arguments leading to sigmoidal flow curves
\cite{Wyart2014}, with the case made above for the generic inadmissibility of
shear-banded steady states, there is every reason to believe that our
observations represent the inherent bulk rheology of very dense suspensions. Of
course, the details of each unsteady flow, particularly in the chaotic regimes,
may depend on the precise sample geometry. In particular it may be influenced
by the finite stress and/or strain-rate gradients imposed by all real
rheometers. Nonetheless, it seems clear that unsteady flow is an intrinsic
element of the rheology of very dense shear-thickening suspensions.

\subsection*{Acknowledgements} We thank Suzanne Fielding, Peter Olmsted, and Romain Mari for discussions.
This work was funded in part by EPSRC Grant EP/J007404. MEC holds a Royal Society Research Professorship. 
BG was funded by a CASE scholarship with Johnson Matthey.
M.W. thanks the Swiss National Science Foundation for support under Grant No. 200021-165509 and the Simons Collaborative Grant ``Crackling the glass problem''. 

\subsection*{Data} 
Data relevant to this work have been archived and can be
accessed \url{http://dx.doi.org/10.7488/ds/1393}.

\bibliography{article_sigmoidal} 

\begin{thebibliography}{46}%
\makeatletter
\providecommand \@ifxundefined [1]{%
 \@ifx{#1\undefined}
}%
\providecommand \@ifnum [1]{%
 \ifnum #1\expandafter \@firstoftwo
 \else \expandafter \@secondoftwo
 \fi
}%
\providecommand \@ifx [1]{%
 \ifx #1\expandafter \@firstoftwo
 \else \expandafter \@secondoftwo
 \fi
}%
\providecommand \natexlab [1]{#1}%
\providecommand \enquote  [1]{``#1''}%
\providecommand \bibnamefont  [1]{#1}%
\providecommand \bibfnamefont [1]{#1}%
\providecommand \citenamefont [1]{#1}%
\providecommand \href@noop [0]{\@secondoftwo}%
\providecommand \href [0]{\begingroup \@sanitize@url \@href}%
\providecommand \@href[1]{\@@startlink{#1}\@@href}%
\providecommand \@@href[1]{\endgroup#1\@@endlink}%
\providecommand \@sanitize@url [0]{\catcode `\\12\catcode `\$12\catcode
  `\&12\catcode `\#12\catcode `\^12\catcode `\_12\catcode `\%12\relax}%
\providecommand \@@startlink[1]{}%
\providecommand \@@endlink[0]{}%
\providecommand \url  [0]{\begingroup\@sanitize@url \@url }%
\providecommand \@url [1]{\endgroup\@href {#1}{\urlprefix }}%
\providecommand \urlprefix  [0]{URL }%
\providecommand \Eprint [0]{\href }%
\providecommand \doibase [0]{http://dx.doi.org/}%
\providecommand \selectlanguage [0]{\@gobble}%
\providecommand \bibinfo  [0]{\@secondoftwo}%
\providecommand \bibfield  [0]{\@secondoftwo}%
\providecommand \translation [1]{[#1]}%
\providecommand \BibitemOpen [0]{}%
\providecommand \bibitemStop [0]{}%
\providecommand \bibitemNoStop [0]{.\EOS\space}%
\providecommand \EOS [0]{\spacefactor3000\relax}%
\providecommand \BibitemShut  [1]{\csname bibitem#1\endcsname}%
\let\auto@bib@innerbib\@empty
\bibitem [{\citenamefont {Olmsted}(2008)}]{olmsted2008}%
  \BibitemOpen
  \bibfield  {author} {\bibinfo {author} {\bibfnamefont {Peter~D.}\
  \bibnamefont {Olmsted}},\ }\bibfield  {title} {\enquote {\bibinfo {title}
  {Perspectives on shear banding in complex fluids},}\ }\href {\doibase
  10.1007/s00397-008-0260-9} {\bibfield  {journal} {\bibinfo  {journal}
  {Rheologica Acta}\ }\textbf {\bibinfo {volume} {47}},\ \bibinfo {pages}
  {283--300} (\bibinfo {year} {2008})}\BibitemShut {NoStop}%
\bibitem [{\citenamefont {Cates}\ and\ \citenamefont
  {Fielding}(2006)}]{cates2006}%
  \BibitemOpen
  \bibfield  {author} {\bibinfo {author} {\bibfnamefont {M.~E.}\ \bibnamefont
  {Cates}}\ and\ \bibinfo {author} {\bibfnamefont {S.~M.}\ \bibnamefont
  {Fielding}},\ }\bibfield  {title} {\enquote {\bibinfo {title} {Rheology of
  giant micelles},}\ }\href {\doibase 10.1080/00018730601082029} {\bibfield
  {journal} {\bibinfo  {journal} {Advances in Physics}\ }\textbf {\bibinfo
  {volume} {55}},\ \bibinfo {pages} {799--879} (\bibinfo {year}
  {2006})}\BibitemShut {NoStop}%
\bibitem [{\citenamefont {Chakrabarti}\ \emph {et~al.}(2004)\citenamefont
  {Chakrabarti}, \citenamefont {Das}, \citenamefont {Dasgupta}, \citenamefont
  {Ramaswamy},\ and\ \citenamefont {Sood}}]{chakrabarti2004}%
  \BibitemOpen
  \bibfield  {author} {\bibinfo {author} {\bibfnamefont {Buddhapriya}\
  \bibnamefont {Chakrabarti}}, \bibinfo {author} {\bibfnamefont {Moumita}\
  \bibnamefont {Das}}, \bibinfo {author} {\bibfnamefont {Chandan}\ \bibnamefont
  {Dasgupta}}, \bibinfo {author} {\bibfnamefont {Sriram}\ \bibnamefont
  {Ramaswamy}}, \ and\ \bibinfo {author} {\bibfnamefont {A.~K.}\ \bibnamefont
  {Sood}},\ }\bibfield  {title} {\enquote {\bibinfo {title} {Spatiotemporal
  {Rheochaos} in {Nematic} {Hydrodynamics}},}\ }\href {\doibase
  10.1103/PhysRevLett.92.055501} {\bibfield  {journal} {\bibinfo  {journal}
  {Physical Review Letters}\ }\textbf {\bibinfo {volume} {92}},\ \bibinfo
  {pages} {055501} (\bibinfo {year} {2004})}\BibitemShut {NoStop}%
\bibitem [{\citenamefont {Einstein}(1905)}]{einstein1905}%
  \BibitemOpen
  \bibfield  {author} {\bibinfo {author} {\bibfnamefont {A.}~\bibnamefont
  {Einstein}},\ }\bibfield  {title} {\enquote {\bibinfo {title} {\"uber die von
  der molekularkinetischen theorie der w\"arme geforderte bewegung von in
  ruhenden {Fl\"ussigkeiten} suspendierten teilchen},}\ }\href
  {http://onlinelibrary.wiley.com/doi/10.1002/andp.19053220806/abstract}
  {\bibfield  {journal} {\bibinfo  {journal} {Ann. Phys.}\ }\textbf {\bibinfo
  {volume} {322}},\ \bibinfo {pages} {549--560} (\bibinfo {year}
  {1905})}\BibitemShut {NoStop}%
\bibitem [{\citenamefont {Batchelor}(1977)}]{batchelor1977}%
  \BibitemOpen
  \bibfield  {author} {\bibinfo {author} {\bibfnamefont {G.~K.}\ \bibnamefont
  {Batchelor}},\ }\bibfield  {title} {\enquote {\bibinfo {title} {The effect of
  {Brownian} motion on the bulk stress in a suspension of spherical
  particles},}\ }\href
  {http://journals.cambridge.org/article_S0022112077001062} {\bibfield
  {journal} {\bibinfo  {journal} {J. Fluid Mech.}\ }\textbf {\bibinfo {volume}
  {83}},\ \bibinfo {pages} {97--117} (\bibinfo {year} {1977})}\BibitemShut
  {NoStop}%
\bibitem [{\citenamefont {G{\"o}tze}(2012)}]{Gotze}%
  \BibitemOpen
  \bibfield  {author} {\bibinfo {author} {\bibfnamefont {W.}~\bibnamefont
  {G{\"o}tze}},\ }\href@noop {} {\emph {\bibinfo {title} {{Complex Dynamics of
  Glass-Forming Liquids: A Mode-Coupling Theory}}}}\ (\bibinfo  {publisher}
  {Oxford University Press},\ \bibinfo {address} {Oxford},\ \bibinfo {year}
  {2012})\BibitemShut {NoStop}%
\bibitem [{\citenamefont {Gadala‐Maria}\ and\ \citenamefont
  {Acrivos}(1980)}]{gadalamaria1980}%
  \BibitemOpen
  \bibfield  {author} {\bibinfo {author} {\bibfnamefont {F.}~\bibnamefont
  {Gadala‐Maria}}\ and\ \bibinfo {author} {\bibfnamefont {Andreas}\
  \bibnamefont {Acrivos}},\ }\bibfield  {title} {\enquote {\bibinfo {title}
  {Shear‐{Induced} {Structure} in a {Concentrated} {Suspension} of {Solid}
  {Spheres}},}\ }\href {\doibase 10.1122/1.549584} {\bibfield  {journal}
  {\bibinfo  {journal} {Journal of Rheology (1978-present)}\ }\textbf {\bibinfo
  {volume} {24}},\ \bibinfo {pages} {799--814} (\bibinfo {year}
  {1980})}\BibitemShut {NoStop}%
\bibitem [{\citenamefont {Wilson}\ and\ \citenamefont
  {Davis}(2000)}]{wilson2000}%
  \BibitemOpen
  \bibfield  {author} {\bibinfo {author} {\bibfnamefont {Helen~J.}\
  \bibnamefont {Wilson}}\ and\ \bibinfo {author} {\bibfnamefont {Robert~H.}\
  \bibnamefont {Davis}},\ }\bibfield  {title} {\enquote {\bibinfo {title} {The
  viscosity of a dilute suspension of rough spheres},}\ }\href
  {http://journals.cambridge.org/article_S0022112000001695} {\bibfield
  {journal} {\bibinfo  {journal} {J. Fluid Mech.}\ }\textbf {\bibinfo {volume}
  {421}},\ \bibinfo {pages} {339--367} (\bibinfo {year} {2000})}\BibitemShut
  {NoStop}%
\bibitem [{\citenamefont {Guy}\ \emph {et~al.}(2015)\citenamefont {Guy},
  \citenamefont {Hermes},\ and\ \citenamefont {Poon}}]{guy2015}%
  \BibitemOpen
  \bibfield  {author} {\bibinfo {author} {\bibfnamefont {BM}~\bibnamefont
  {Guy}}, \bibinfo {author} {\bibfnamefont {M}~\bibnamefont {Hermes}}, \ and\
  \bibinfo {author} {\bibfnamefont {WCK}\ \bibnamefont {Poon}},\ }\bibfield
  {title} {\enquote {\bibinfo {title} {Towards a unified description of the
  rheology of hard-particle suspensions},}\ }\href@noop {} {\bibfield
  {journal} {\bibinfo  {journal} {Phys. Rev. Lett.}\ }\textbf {\bibinfo
  {volume} {115}},\ \bibinfo {pages} {088304} (\bibinfo {year}
  {2015})}\BibitemShut {NoStop}%
\bibitem [{\citenamefont {Lin}\ \emph {et~al.}(2015)\citenamefont {Lin},
  \citenamefont {Guy}, \citenamefont {Hermes}, \citenamefont {Ness},
  \citenamefont {Sun}, \citenamefont {Poon},\ and\ \citenamefont
  {Cohen}}]{lin2015}%
  \BibitemOpen
  \bibfield  {author} {\bibinfo {author} {\bibfnamefont {Neil~YC}\ \bibnamefont
  {Lin}}, \bibinfo {author} {\bibfnamefont {Ben~M}\ \bibnamefont {Guy}},
  \bibinfo {author} {\bibfnamefont {Michiel}\ \bibnamefont {Hermes}}, \bibinfo
  {author} {\bibfnamefont {Chris}\ \bibnamefont {Ness}}, \bibinfo {author}
  {\bibfnamefont {Jin}\ \bibnamefont {Sun}}, \bibinfo {author} {\bibfnamefont
  {Wilson~CK}\ \bibnamefont {Poon}}, \ and\ \bibinfo {author} {\bibfnamefont
  {Itai}\ \bibnamefont {Cohen}},\ }\bibfield  {title} {\enquote {\bibinfo
  {title} {Hydrodynamic and contact contributions to continuous shear
  thickening in colloidal suspensions},}\ }\href@noop {} {\bibfield  {journal}
  {\bibinfo  {journal} {Physical review letters}\ }\textbf {\bibinfo {volume}
  {115}},\ \bibinfo {pages} {228304} (\bibinfo {year} {2015})}\BibitemShut
  {NoStop}%
\bibitem [{\citenamefont {Seto}\ \emph {et~al.}(2013)\citenamefont {Seto},
  \citenamefont {Mari}, \citenamefont {Morris},\ and\ \citenamefont
  {Denn}}]{Seto2013}%
  \BibitemOpen
  \bibfield  {author} {\bibinfo {author} {\bibfnamefont {Ryohei}\ \bibnamefont
  {Seto}}, \bibinfo {author} {\bibfnamefont {Romain}\ \bibnamefont {Mari}},
  \bibinfo {author} {\bibfnamefont {Jeffrey~F.}\ \bibnamefont {Morris}}, \ and\
  \bibinfo {author} {\bibfnamefont {Morton~M.}\ \bibnamefont {Denn}},\
  }\bibfield  {title} {\enquote {\bibinfo {title} {{Discontinuous Shear
  Thickening of Frictional Hard-Sphere Suspensions}},}\ }\href {\doibase
  10.1103/PhysRevLett.111.218301} {\bibfield  {journal} {\bibinfo  {journal}
  {Phys. Rev. Lett.}\ }\textbf {\bibinfo {volume} {111}},\ \bibinfo {pages}
  {218301} (\bibinfo {year} {2013})}\BibitemShut {NoStop}%
\bibitem [{\citenamefont {Mari}\ \emph {et~al.}(2014)\citenamefont {Mari},
  \citenamefont {Seto}, \citenamefont {Morris},\ and\ \citenamefont
  {Denn}}]{Mari2014}%
  \BibitemOpen
  \bibfield  {author} {\bibinfo {author} {\bibfnamefont {Romain}\ \bibnamefont
  {Mari}}, \bibinfo {author} {\bibfnamefont {Ryohei}\ \bibnamefont {Seto}},
  \bibinfo {author} {\bibfnamefont {Jeffrey~F}\ \bibnamefont {Morris}}, \ and\
  \bibinfo {author} {\bibfnamefont {Morton~M}\ \bibnamefont {Denn}},\
  }\bibfield  {title} {\enquote {\bibinfo {title} {Shear thickening,
  frictionless and frictional rheologies in non-brownian suspensions},}\
  }\href@noop {} {\bibfield  {journal} {\bibinfo  {journal} {J. Rheol.}\
  }\textbf {\bibinfo {volume} {58}},\ \bibinfo {pages} {1693--1724} (\bibinfo
  {year} {2014})}\BibitemShut {NoStop}%
\bibitem [{\citenamefont {Mari}\ \emph {et~al.}(2015)\citenamefont {Mari},
  \citenamefont {Seto}, \citenamefont {Morris},\ and\ \citenamefont
  {Denn}}]{Mari2015}%
  \BibitemOpen
  \bibfield  {author} {\bibinfo {author} {\bibfnamefont {Romain}\ \bibnamefont
  {Mari}}, \bibinfo {author} {\bibfnamefont {Ryohei}\ \bibnamefont {Seto}},
  \bibinfo {author} {\bibfnamefont {Jeffrey~F.}\ \bibnamefont {Morris}}, \ and\
  \bibinfo {author} {\bibfnamefont {Morton~M.}\ \bibnamefont {Denn}},\
  }\bibfield  {title} {\enquote {\bibinfo {title} {Nonmonotonic flow curves of
  shear thickening suspensions},}\ }\href {\doibase 10.1103/PhysRevE.91.052302}
  {\bibfield  {journal} {\bibinfo  {journal} {Phys. Rev. E}\ }\textbf {\bibinfo
  {volume} {91}},\ \bibinfo {pages} {052302} (\bibinfo {year}
  {2015})}\BibitemShut {NoStop}%
\bibitem [{\citenamefont {Ness}\ and\ \citenamefont {Sun}(2016)}]{ness2016}%
  \BibitemOpen
  \bibfield  {author} {\bibinfo {author} {\bibfnamefont {Christopher}\
  \bibnamefont {Ness}}\ and\ \bibinfo {author} {\bibfnamefont {Jin}\
  \bibnamefont {Sun}},\ }\bibfield  {title} {\enquote {\bibinfo {title} {Shear
  thickening regimes of dense non-brownian suspensions},}\ }\href@noop {}
  {\bibfield  {journal} {\bibinfo  {journal} {Soft matter}\ }\textbf {\bibinfo
  {volume} {12}},\ \bibinfo {pages} {914--924} (\bibinfo {year}
  {2016})}\BibitemShut {NoStop}%
\bibitem [{\citenamefont {Wyart}\ and\ \citenamefont
  {Cates}(2014)}]{Wyart2014}%
  \BibitemOpen
  \bibfield  {author} {\bibinfo {author} {\bibfnamefont {M.}~\bibnamefont
  {Wyart}}\ and\ \bibinfo {author} {\bibfnamefont {M.~E.}\ \bibnamefont
  {Cates}},\ }\bibfield  {title} {\enquote {\bibinfo {title} {{Discontinuous
  Shear Thickening without Inertia in Dense Non-Brownian Suspensions}},}\
  }\href {\doibase 10.1103/PhysRevLett.112.098302} {\bibfield  {journal}
  {\bibinfo  {journal} {Phys. Rev. Lett.}\ }\textbf {\bibinfo {volume} {112}},\
  \bibinfo {pages} {098302} (\bibinfo {year} {2014})}\BibitemShut {NoStop}%
\bibitem [{\citenamefont {Melrose}\ and\ \citenamefont
  {Ball}(1995)}]{melrose1995}%
  \BibitemOpen
  \bibfield  {author} {\bibinfo {author} {\bibfnamefont {J.~R}\ \bibnamefont
  {Melrose}}\ and\ \bibinfo {author} {\bibfnamefont {R.~C}\ \bibnamefont
  {Ball}},\ }\bibfield  {title} {\enquote {\bibinfo {title} {The {Pathological}
  {Behaviour} of {Sheared} {Hard} {Spheres} with {Hydrodynamic}
  {Interactions}},}\ }\href@noop {} {\bibfield  {journal} {\bibinfo  {journal}
  {Europhys. Lett.}\ }\textbf {\bibinfo {volume} {32}},\ \bibinfo {pages}
  {535--540} (\bibinfo {year} {1995})}\BibitemShut {NoStop}%
\bibitem [{\citenamefont {Melrose}\ \emph {et~al.}(1996)\citenamefont
  {Melrose}, \citenamefont {van Vliet},\ and\ \citenamefont
  {Ball}}]{melrose1996}%
  \BibitemOpen
  \bibfield  {author} {\bibinfo {author} {\bibfnamefont {J.~R.}\ \bibnamefont
  {Melrose}}, \bibinfo {author} {\bibfnamefont {J.~H.}\ \bibnamefont {van
  Vliet}}, \ and\ \bibinfo {author} {\bibfnamefont {R.~C.}\ \bibnamefont
  {Ball}},\ }\bibfield  {title} {\enquote {\bibinfo {title} {Continuous {Shear}
  {Thickening} and {Colloid} {Surfaces}},}\ }\href {\doibase
  10.1103/PhysRevLett.77.4660} {\bibfield  {journal} {\bibinfo  {journal}
  {Phys. Rev. Lett.}\ }\textbf {\bibinfo {volume} {77}},\ \bibinfo {pages}
  {4660--4663} (\bibinfo {year} {1996})}\BibitemShut {NoStop}%
\bibitem [{\citenamefont {Morris}(2009)}]{morris2009}%
  \BibitemOpen
  \bibfield  {author} {\bibinfo {author} {\bibfnamefont {Jeffrey~F.}\
  \bibnamefont {Morris}},\ }\bibfield  {title} {\enquote {\bibinfo {title} {A
  review of microstructure in concentrated suspensions and its implications for
  rheology and bulk flow},}\ }\href {\doibase 10.1007/s00397-009-0352-1}
  {\bibfield  {journal} {\bibinfo  {journal} {Rheologica Acta}\ }\textbf
  {\bibinfo {volume} {48}},\ \bibinfo {pages} {909--923} (\bibinfo {year}
  {2009})}\BibitemShut {NoStop}%
\bibitem [{\citenamefont {Fernandez}\ \emph {et~al.}(2013)\citenamefont
  {Fernandez}, \citenamefont {Mani}, \citenamefont {Rinaldi}, \citenamefont
  {Kadau}, \citenamefont {Mosquet}, \citenamefont {Lombois-Burger},
  \citenamefont {Cayer-Barrioz}, \citenamefont {Herrmann}, \citenamefont
  {Spencer},\ and\ \citenamefont {Isa}}]{Fernandez2013}%
  \BibitemOpen
  \bibfield  {author} {\bibinfo {author} {\bibfnamefont {Nicolas}\ \bibnamefont
  {Fernandez}}, \bibinfo {author} {\bibfnamefont {Roman}\ \bibnamefont {Mani}},
  \bibinfo {author} {\bibfnamefont {David}\ \bibnamefont {Rinaldi}}, \bibinfo
  {author} {\bibfnamefont {Dirk}\ \bibnamefont {Kadau}}, \bibinfo {author}
  {\bibfnamefont {Martin}\ \bibnamefont {Mosquet}}, \bibinfo {author}
  {\bibfnamefont {H\'{e}l\`{e}ne}\ \bibnamefont {Lombois-Burger}}, \bibinfo
  {author} {\bibfnamefont {Juliette}\ \bibnamefont {Cayer-Barrioz}}, \bibinfo
  {author} {\bibfnamefont {Hans~J.}\ \bibnamefont {Herrmann}}, \bibinfo
  {author} {\bibfnamefont {Nicholas~D.}\ \bibnamefont {Spencer}}, \ and\
  \bibinfo {author} {\bibfnamefont {Lucio}\ \bibnamefont {Isa}},\ }\bibfield
  {title} {\enquote {\bibinfo {title} {{Microscopic Mechanism for Shear
  Thickening of Non-Brownian Suspensions}},}\ }\href {\doibase
  10.1103/PhysRevLett.111.108301} {\bibfield  {journal} {\bibinfo  {journal}
  {Phys. Rev. Lett.}\ }\textbf {\bibinfo {volume} {111}},\ \bibinfo {pages}
  {108301} (\bibinfo {year} {2013})}\BibitemShut {NoStop}%
\bibitem [{\citenamefont {Bender}(1996)}]{Bender1996}%
  \BibitemOpen
  \bibfield  {author} {\bibinfo {author} {\bibfnamefont {Jonathan}\
  \bibnamefont {Bender}},\ }\bibfield  {title} {\enquote {\bibinfo {title}
  {{Reversible shear thickening in monodisperse and bidisperse colloidal
  dispersions}},}\ }\href {\doibase 10.1122/1.550767} {\bibfield  {journal}
  {\bibinfo  {journal} {J. Rheol.}\ }\textbf {\bibinfo {volume} {40}},\
  \bibinfo {pages} {899} (\bibinfo {year} {1996})}\BibitemShut {NoStop}%
\bibitem [{Note1()}]{Note1}%
  \BibitemOpen
  \bibinfo {note} {Note that the stress measured on the top plate is a
  combination of the of the first and second normal stress differences and not
  a measurement of the particles pressure as measured by \cite
  {Boyer2011}.}\BibitemShut {Stop}%
\bibitem [{\citenamefont {Grob}\ \emph {et~al.}(2015)\citenamefont {Grob},
  \citenamefont {Zippelius},\ and\ \citenamefont {Heussinger}}]{grob2015}%
  \BibitemOpen
  \bibfield  {author} {\bibinfo {author} {\bibfnamefont {Matthias}\
  \bibnamefont {Grob}}, \bibinfo {author} {\bibfnamefont {Annette}\
  \bibnamefont {Zippelius}}, \ and\ \bibinfo {author} {\bibfnamefont {Claus}\
  \bibnamefont {Heussinger}},\ }\bibfield  {title} {\enquote {\bibinfo {title}
  {Rheo-chaos of frictional grains},}\ }\href@noop {} {\bibfield  {journal}
  {\bibinfo  {journal} {arXiv preprint arXiv:1507.07421}\ } (\bibinfo {year}
  {2015})}\BibitemShut {NoStop}%
\bibitem [{\citenamefont {Boyer}\ \emph {et~al.}(2011)\citenamefont {Boyer},
  \citenamefont {Guazzelli},\ and\ \citenamefont {Pouliquen}}]{Boyer2011}%
  \BibitemOpen
  \bibfield  {author} {\bibinfo {author} {\bibfnamefont {Fran\c{c}ois}\
  \bibnamefont {Boyer}}, \bibinfo {author} {\bibfnamefont {\'{E}lisabeth}\
  \bibnamefont {Guazzelli}}, \ and\ \bibinfo {author} {\bibfnamefont {Olivier}\
  \bibnamefont {Pouliquen}},\ }\bibfield  {title} {\enquote {\bibinfo {title}
  {{Unifying Suspension and Granular Rheology}},}\ }\href {\doibase
  10.1103/PhysRevLett.107.188301} {\bibfield  {journal} {\bibinfo  {journal}
  {Phys. Rev. Lett.}\ }\textbf {\bibinfo {volume} {107}},\ \bibinfo {pages}
  {188301} (\bibinfo {year} {2011})}\BibitemShut {NoStop}%
\bibitem [{\citenamefont {Trulsson}\ \emph {et~al.}(2012)\citenamefont
  {Trulsson}, \citenamefont {Andreotti},\ and\ \citenamefont
  {Claudin}}]{trulsson2012}%
  \BibitemOpen
  \bibfield  {author} {\bibinfo {author} {\bibfnamefont {Martin}\ \bibnamefont
  {Trulsson}}, \bibinfo {author} {\bibfnamefont {Bruno}\ \bibnamefont
  {Andreotti}}, \ and\ \bibinfo {author} {\bibfnamefont {Philippe}\
  \bibnamefont {Claudin}},\ }\bibfield  {title} {\enquote {\bibinfo {title}
  {Transition from the viscous to inertial regime in dense suspensions},}\
  }\href@noop {} {\bibfield  {journal} {\bibinfo  {journal} {Phys. Rev. Lett.}\
  }\textbf {\bibinfo {volume} {109}},\ \bibinfo {pages} {118305} (\bibinfo
  {year} {2012})}\BibitemShut {NoStop}%
\bibitem [{\citenamefont {Lemaître}\ \emph {et~al.}(2009)\citenamefont
  {Lemaître}, \citenamefont {Roux},\ and\ \citenamefont
  {Chevoir}}]{lemaitre2009}%
  \BibitemOpen
  \bibfield  {author} {\bibinfo {author} {\bibfnamefont {Anaël}\ \bibnamefont
  {Lemaître}}, \bibinfo {author} {\bibfnamefont {Jean-Noël}\ \bibnamefont
  {Roux}}, \ and\ \bibinfo {author} {\bibfnamefont {François}\ \bibnamefont
  {Chevoir}},\ }\bibfield  {title} {\enquote {\bibinfo {title} {What do dry
  granular flows tell us about dense non-{Brownian} suspension rheology?}}\
  }\href {\doibase 10.1007/s00397-009-0379-3} {\bibfield  {journal} {\bibinfo
  {journal} {Rheologica Acta}\ }\textbf {\bibinfo {volume} {48}},\ \bibinfo
  {pages} {925--942} (\bibinfo {year} {2009})}\BibitemShut {NoStop}%
\bibitem [{\citenamefont {Lerner}\ \emph {et~al.}(2012)\citenamefont {Lerner},
  \citenamefont {D{\"u}ring},\ and\ \citenamefont {Wyart}}]{lerner2012}%
  \BibitemOpen
  \bibfield  {author} {\bibinfo {author} {\bibfnamefont {Edan}\ \bibnamefont
  {Lerner}}, \bibinfo {author} {\bibfnamefont {Gustavo}\ \bibnamefont
  {D{\"u}ring}}, \ and\ \bibinfo {author} {\bibfnamefont {Matthieu}\
  \bibnamefont {Wyart}},\ }\bibfield  {title} {\enquote {\bibinfo {title} {A
  unified framework for non-brownian suspension flows and soft amorphous
  solids},}\ }\href@noop {} {\bibfield  {journal} {\bibinfo  {journal}
  {Proceedings of the National Academy of Sciences}\ }\textbf {\bibinfo
  {volume} {109}},\ \bibinfo {pages} {4798--4803} (\bibinfo {year}
  {2012})}\BibitemShut {NoStop}%
\bibitem [{\citenamefont {Andreotti}\ \emph {et~al.}(2012)\citenamefont
  {Andreotti}, \citenamefont {Barrat},\ and\ \citenamefont
  {Heussinger}}]{andreotti2012}%
  \BibitemOpen
  \bibfield  {author} {\bibinfo {author} {\bibfnamefont {Bruno}\ \bibnamefont
  {Andreotti}}, \bibinfo {author} {\bibfnamefont {Jean-Louis}\ \bibnamefont
  {Barrat}}, \ and\ \bibinfo {author} {\bibfnamefont {Claus}\ \bibnamefont
  {Heussinger}},\ }\bibfield  {title} {\enquote {\bibinfo {title} {Shear flow
  of non-brownian suspensions close to jamming},}\ }\href@noop {} {\bibfield
  {journal} {\bibinfo  {journal} {Physical review letters}\ }\textbf {\bibinfo
  {volume} {109}},\ \bibinfo {pages} {105901} (\bibinfo {year}
  {2012})}\BibitemShut {NoStop}%
\bibitem [{\citenamefont {DeGiuli}\ \emph {et~al.}(2015)\citenamefont
  {DeGiuli}, \citenamefont {D\"uring}, \citenamefont {Lerner},\ and\
  \citenamefont {Wyart}}]{degiuli2015}%
  \BibitemOpen
  \bibfield  {author} {\bibinfo {author} {\bibfnamefont {E.}~\bibnamefont
  {DeGiuli}}, \bibinfo {author} {\bibfnamefont {G.}~\bibnamefont {D\"uring}},
  \bibinfo {author} {\bibfnamefont {E.}~\bibnamefont {Lerner}}, \ and\ \bibinfo
  {author} {\bibfnamefont {M.}~\bibnamefont {Wyart}},\ }\bibfield  {title}
  {\enquote {\bibinfo {title} {Unified theory of inertial granular flows and
  non-brownian suspensions},}\ }\href {\doibase 10.1103/PhysRevE.91.062206}
  {\bibfield  {journal} {\bibinfo  {journal} {Phys. Rev. E}\ }\textbf {\bibinfo
  {volume} {91}},\ \bibinfo {pages} {062206} (\bibinfo {year}
  {2015})}\BibitemShut {NoStop}%
\bibitem [{\citenamefont {Bashkirtseva}\ \emph {et~al.}(2009)\citenamefont
  {Bashkirtseva}, \citenamefont {Zubarev}, \citenamefont {Iskakova},\ and\
  \citenamefont {Ryashko}}]{bashkirtseva2009}%
  \BibitemOpen
  \bibfield  {author} {\bibinfo {author} {\bibfnamefont {I.~A.}\ \bibnamefont
  {Bashkirtseva}}, \bibinfo {author} {\bibfnamefont {A.~Yu.}\ \bibnamefont
  {Zubarev}}, \bibinfo {author} {\bibfnamefont {L.~Yu.}\ \bibnamefont
  {Iskakova}}, \ and\ \bibinfo {author} {\bibfnamefont {L.~B.}\ \bibnamefont
  {Ryashko}},\ }\bibfield  {title} {\enquote {\bibinfo {title} {{On rheophysics
  of high-concentrated suspensions}},}\ }\href {\doibase
  10.1134/S1061933X09040024} {\bibfield  {journal} {\bibinfo  {journal}
  {Colloid Journal}\ }\textbf {\bibinfo {volume} {71}},\ \bibinfo {pages}
  {446--454} (\bibinfo {year} {2009})}\BibitemShut {NoStop}%
\bibitem [{\citenamefont {Laun}(1994)}]{Laun1994}%
  \BibitemOpen
  \bibfield  {author} {\bibinfo {author} {\bibfnamefont {H.M.}\ \bibnamefont
  {Laun}},\ }\bibfield  {title} {\enquote {\bibinfo {title} {{Normal stresses
  in extremely shear thickening polymer dispersions}},}\ }\href {\doibase
  10.1016/0377-0257(94)80016-2} {\bibfield  {journal} {\bibinfo  {journal} {J.
  Non-Newt. Fluid Mech.}\ }\textbf {\bibinfo {volume} {54}},\ \bibinfo {pages}
  {87--108} (\bibinfo {year} {1994})}\BibitemShut {NoStop}%
\bibitem [{\citenamefont {Cates}\ \emph {et~al.}(2005)\citenamefont {Cates},
  \citenamefont {Haw},\ and\ \citenamefont {Holmes}}]{cates2005}%
  \BibitemOpen
  \bibfield  {author} {\bibinfo {author} {\bibfnamefont {M.~E.}\ \bibnamefont
  {Cates}}, \bibinfo {author} {\bibfnamefont {M.~D.}\ \bibnamefont {Haw}}, \
  and\ \bibinfo {author} {\bibfnamefont {C.~B.}\ \bibnamefont {Holmes}},\
  }\bibfield  {title} {\enquote {\bibinfo {title} {Dilatancy, jamming, and the
  physics of granulation},}\ }\href {\doibase 10.1088/0953-8984/17/24/010}
  {\bibfield  {journal} {\bibinfo  {journal} {Journal of Physics: Condensed
  Matter}\ }\textbf {\bibinfo {volume} {17}},\ \bibinfo {pages} {S2517}
  (\bibinfo {year} {2005})}\BibitemShut {NoStop}%
\bibitem [{\citenamefont {Pan}\ \emph {et~al.}(2015)\citenamefont {Pan},
  \citenamefont {de~Cagny}, \citenamefont {Weber},\ and\ \citenamefont
  {Bonn}}]{pan2015}%
  \BibitemOpen
  \bibfield  {author} {\bibinfo {author} {\bibfnamefont {Zhongcheng}\
  \bibnamefont {Pan}}, \bibinfo {author} {\bibfnamefont {Henri}\ \bibnamefont
  {de~Cagny}}, \bibinfo {author} {\bibfnamefont {Bart}\ \bibnamefont {Weber}},
  \ and\ \bibinfo {author} {\bibfnamefont {Daniel}\ \bibnamefont {Bonn}},\
  }\bibfield  {title} {\enquote {\bibinfo {title} {$\mathsf{S}$-shaped flow
  curves of shear thickening suspensions: {Direct} observation of frictional
  rheology},}\ }\href {\doibase 10.1103/PhysRevE.92.032202} {\bibfield
  {journal} {\bibinfo  {journal} {Phys. Rev. E}\ }\textbf {\bibinfo {volume}
  {92}},\ \bibinfo {pages} {032202} (\bibinfo {year} {2015})}\BibitemShut
  {NoStop}%
\bibitem [{\citenamefont {Fielding}\ and\ \citenamefont
  {Olmsted}(2004)}]{fielding2004}%
  \BibitemOpen
  \bibfield  {author} {\bibinfo {author} {\bibfnamefont {S.~M.}\ \bibnamefont
  {Fielding}}\ and\ \bibinfo {author} {\bibfnamefont {P.~D.}\ \bibnamefont
  {Olmsted}},\ }\bibfield  {title} {\enquote {\bibinfo {title} {Spatiotemporal
  {Oscillations} and {Rheochaos} in a {Simple} {Model} of {Shear} {Banding}},}\
  }\href {\doibase 10.1103/PhysRevLett.92.084502} {\bibfield  {journal}
  {\bibinfo  {journal} {Physical Review Letters}\ }\textbf {\bibinfo {volume}
  {92}},\ \bibinfo {pages} {084502} (\bibinfo {year} {2004})}\BibitemShut
  {NoStop}%
\bibitem [{\citenamefont {Andreotti}\ \emph {et~al.}(2013)\citenamefont
  {Andreotti}, \citenamefont {Forterre},\ and\ \citenamefont
  {Pouliquen}}]{andreotti2013}%
  \BibitemOpen
  \bibfield  {author} {\bibinfo {author} {\bibfnamefont {Bruno}\ \bibnamefont
  {Andreotti}}, \bibinfo {author} {\bibfnamefont {Yoël}\ \bibnamefont
  {Forterre}}, \ and\ \bibinfo {author} {\bibfnamefont {Olivier}\ \bibnamefont
  {Pouliquen}},\ }\href@noop {} {\emph {\bibinfo {title} {Granular {Media}:
  {Between} {Fluid} and {Solid}}}}\ (\bibinfo  {publisher} {Cambridge
  University Press},\ \bibinfo {year} {2013})\BibitemShut {NoStop}%
\bibitem [{Note2()}]{Note2}%
  \BibitemOpen
  \bibinfo {note} {Recall that the normal-normal component of the particle
  stress is (minus) the diagonal component of the stress tensor. Thus if the
  $z$ direction is the direction normal to the interface the normal-normal
  component is the $z$-$z$ component of the stress tensor.}\BibitemShut {Stop}%
\bibitem [{\citenamefont {Fall}\ \emph {et~al.}(2015)\citenamefont {Fall},
  \citenamefont {Bertrand}, \citenamefont {Hautemayou}, \citenamefont
  {Mezière}, \citenamefont {Moucheront}, \citenamefont {Lemaître},\ and\
  \citenamefont {Ovarlez}}]{Fall2015}%
  \BibitemOpen
  \bibfield  {author} {\bibinfo {author} {\bibfnamefont {A.}~\bibnamefont
  {Fall}}, \bibinfo {author} {\bibfnamefont {F.}~\bibnamefont {Bertrand}},
  \bibinfo {author} {\bibfnamefont {D.}~\bibnamefont {Hautemayou}}, \bibinfo
  {author} {\bibfnamefont {C.}~\bibnamefont {Mezière}}, \bibinfo {author}
  {\bibfnamefont {P.}~\bibnamefont {Moucheront}}, \bibinfo {author}
  {\bibfnamefont {A.}~\bibnamefont {Lemaître}}, \ and\ \bibinfo {author}
  {\bibfnamefont {G.}~\bibnamefont {Ovarlez}},\ }\bibfield  {title} {\enquote
  {\bibinfo {title} {Macroscopic {Discontinuous} {Shear} {Thickening} versus
  {Local} {Shear} {Jamming} in {Cornstarch}},}\ }\href@noop {} {\bibfield
  {journal} {\bibinfo  {journal} {Phys. Rev. Lett.}\ }\textbf {\bibinfo
  {volume} {114}},\ \bibinfo {pages} {098301} (\bibinfo {year}
  {2015})}\BibitemShut {NoStop}%
\bibitem [{\citenamefont {Adams}\ \emph {et~al.}(2011)\citenamefont {Adams},
  \citenamefont {Fielding},\ and\ \citenamefont {Olmsted}}]{adams2011}%
  \BibitemOpen
  \bibfield  {author} {\bibinfo {author} {\bibfnamefont {J.~M.}\ \bibnamefont
  {Adams}}, \bibinfo {author} {\bibfnamefont {S.~M.}\ \bibnamefont {Fielding}},
  \ and\ \bibinfo {author} {\bibfnamefont {P.~D.}\ \bibnamefont {Olmsted}},\
  }\bibfield  {title} {\enquote {\bibinfo {title} {Transient shear banding in
  entangled polymers: {A} study using the {Rolie}-{Poly} model},}\ }\href
  {\doibase 10.1122/1.3610169} {\bibfield  {journal} {\bibinfo  {journal}
  {Journal of Rheology (1978-present)}\ }\textbf {\bibinfo {volume} {55}},\
  \bibinfo {pages} {1007--1032} (\bibinfo {year} {2011})}\BibitemShut {NoStop}%
\bibitem [{\citenamefont {Skorski}\ and\ \citenamefont
  {Olmsted}(2011)}]{skorski2011}%
  \BibitemOpen
  \bibfield  {author} {\bibinfo {author} {\bibfnamefont {S.}~\bibnamefont
  {Skorski}}\ and\ \bibinfo {author} {\bibfnamefont {P.~D.}\ \bibnamefont
  {Olmsted}},\ }\bibfield  {title} {\enquote {\bibinfo {title} {Loss of
  solutions in shear banding fluids driven by second normal stress
  differences},}\ }\href {\doibase 10.1122/1.3621521} {\bibfield  {journal}
  {\bibinfo  {journal} {Journal of Rheology (1978-present)}\ }\textbf {\bibinfo
  {volume} {55}},\ \bibinfo {pages} {1219--1246} (\bibinfo {year}
  {2011})}\BibitemShut {NoStop}%
\bibitem [{\citenamefont {Helfand}\ and\ \citenamefont
  {Fredrickson}(1989)}]{helfand1989}%
  \BibitemOpen
  \bibfield  {author} {\bibinfo {author} {\bibfnamefont {Eugene}\ \bibnamefont
  {Helfand}}\ and\ \bibinfo {author} {\bibfnamefont {Glenn~H.}\ \bibnamefont
  {Fredrickson}},\ }\bibfield  {title} {\enquote {\bibinfo {title} {Large
  fluctuations in polymer solutions under shear},}\ }\href {\doibase
  10.1103/PhysRevLett.62.2468} {\bibfield  {journal} {\bibinfo  {journal}
  {Phys. Rev. Lett.}\ }\textbf {\bibinfo {volume} {62}},\ \bibinfo {pages}
  {2468--2471} (\bibinfo {year} {1989})}\BibitemShut {NoStop}%
\bibitem [{\citenamefont {von Kann}\ \emph {et~al.}(2011)\citenamefont {von
  Kann}, \citenamefont {Snoeijer}, \citenamefont {Lohse},\ and\ \citenamefont
  {van~der Meer}}]{kann2011}%
  \BibitemOpen
  \bibfield  {author} {\bibinfo {author} {\bibfnamefont {Stefan}\ \bibnamefont
  {von Kann}}, \bibinfo {author} {\bibfnamefont {Jacco~H.}\ \bibnamefont
  {Snoeijer}}, \bibinfo {author} {\bibfnamefont {Detlef}\ \bibnamefont
  {Lohse}}, \ and\ \bibinfo {author} {\bibfnamefont {Devaraj}\ \bibnamefont
  {van~der Meer}},\ }\bibfield  {title} {\enquote {\bibinfo {title}
  {Nonmonotonic settling of a sphere in a cornstarch suspension},}\ }\href
  {\doibase 10.1103/PhysRevE.84.060401} {\bibfield  {journal} {\bibinfo
  {journal} {Physical Review E}\ }\textbf {\bibinfo {volume} {84}},\ \bibinfo
  {pages} {060401} (\bibinfo {year} {2011})}\BibitemShut {NoStop}%
\bibitem [{\citenamefont {Ganapathy}\ and\ \citenamefont
  {Sood}(2006)}]{Ganapathy2006}%
  \BibitemOpen
  \bibfield  {author} {\bibinfo {author} {\bibfnamefont {Rajesh}\ \bibnamefont
  {Ganapathy}}\ and\ \bibinfo {author} {\bibfnamefont {A.~K.}\ \bibnamefont
  {Sood}},\ }\bibfield  {title} {\enquote {\bibinfo {title} {Intermittency
  {Route} to {Rheochaos} in {Wormlike} {Micelles} with {Flow}-{Concentration}
  {Coupling}},}\ }\href {\doibase 10.1103/PhysRevLett.96.108301} {\bibfield
  {journal} {\bibinfo  {journal} {Physical Review Letters}\ }\textbf {\bibinfo
  {volume} {96}},\ \bibinfo {pages} {108301} (\bibinfo {year}
  {2006})}\BibitemShut {NoStop}%
\bibitem [{\citenamefont {Larsen}\ \emph {et~al.}(2014)\citenamefont {Larsen},
  \citenamefont {Kim}, \citenamefont {Zukoski},\ and\ \citenamefont
  {Weitz}}]{larsen2014}%
  \BibitemOpen
  \bibfield  {author} {\bibinfo {author} {\bibfnamefont {Ryan~J.}\ \bibnamefont
  {Larsen}}, \bibinfo {author} {\bibfnamefont {Jin-Woong}\ \bibnamefont {Kim}},
  \bibinfo {author} {\bibfnamefont {Charles~F.}\ \bibnamefont {Zukoski}}, \
  and\ \bibinfo {author} {\bibfnamefont {David~a.}\ \bibnamefont {Weitz}},\
  }\bibfield  {title} {\enquote {\bibinfo {title} {{Fluctuations in flow
  produced by competition between apparent wall slip and dilatancy}},}\ }\href
  {\doibase 10.1007/s00397-014-0764-4} {\bibfield  {journal} {\bibinfo
  {journal} {Rheol. Acta}\ }\textbf {\bibinfo {volume} {53}},\ \bibinfo {pages}
  {333--347} (\bibinfo {year} {2014})}\BibitemShut {NoStop}%
\bibitem [{\citenamefont {Deboeuf}\ \emph {et~al.}(2009)\citenamefont
  {Deboeuf}, \citenamefont {Gauthier}, \citenamefont {Martin}, \citenamefont
  {Yurkovetsky},\ and\ \citenamefont {Morris}}]{deboeuf2009}%
  \BibitemOpen
  \bibfield  {author} {\bibinfo {author} {\bibfnamefont {Angélique}\
  \bibnamefont {Deboeuf}}, \bibinfo {author} {\bibfnamefont {Georges}\
  \bibnamefont {Gauthier}}, \bibinfo {author} {\bibfnamefont {Jérôme}\
  \bibnamefont {Martin}}, \bibinfo {author} {\bibfnamefont {Yevgeny}\
  \bibnamefont {Yurkovetsky}}, \ and\ \bibinfo {author} {\bibfnamefont
  {Jeffrey~F.}\ \bibnamefont {Morris}},\ }\bibfield  {title} {\enquote
  {\bibinfo {title} {Particle {Pressure} in a {Sheared} {Suspension}: {A}
  {Bridge} from {Osmosis} to {Granular} {Dilatancy}},}\ }\href {\doibase
  10.1103/PhysRevLett.102.108301} {\bibfield  {journal} {\bibinfo  {journal}
  {Physical Review Letters}\ }\textbf {\bibinfo {volume} {102}},\ \bibinfo
  {pages} {108301} (\bibinfo {year} {2009})}\BibitemShut {NoStop}%
\bibitem [{\citenamefont {Garland}\ \emph {et~al.}(2013)\citenamefont
  {Garland}, \citenamefont {Gauthier}, \citenamefont {Martin},\ and\
  \citenamefont {~}}]{Garland2013}%
  \BibitemOpen
  \bibfield  {author} {\bibinfo {author} {\bibfnamefont {S.}~\bibnamefont
  {Garland}}, \bibinfo {author} {\bibfnamefont {G.}~\bibnamefont {Gauthier}},
  \bibinfo {author} {\bibfnamefont {J.}~\bibnamefont {Martin}}, \ and\ \bibinfo
  {author} {\bibfnamefont {J.~F.}\ \bibnamefont {~}},\ }\bibfield  {title}
  {\enquote {\bibinfo {title} {{Normal stress measurements in sheared
  non-Brownian suspensions}},}\ }\href {\doibase 10.1122/1.4758001} {\bibfield
  {journal} {\bibinfo  {journal} {J. Rheol.}\ }\textbf {\bibinfo {volume}
  {57}},\ \bibinfo {pages} {71} (\bibinfo {year} {2013})}\BibitemShut {NoStop}%
\bibitem [{\citenamefont {Cwalina}\ and\ \citenamefont
  {Wagner}(2014)}]{Cwalina2014}%
  \BibitemOpen
  \bibfield  {author} {\bibinfo {author} {\bibfnamefont {Colin~D}\ \bibnamefont
  {Cwalina}}\ and\ \bibinfo {author} {\bibfnamefont {Norman~J}\ \bibnamefont
  {Wagner}},\ }\bibfield  {title} {\enquote {\bibinfo {title} {{Material
  properties of the shear-thickened state in concentrated near hard-sphere
  colloidal dispersions}},}\ }\href@noop {} {\bibfield  {journal} {\bibinfo
  {journal} {J. Rheol.}\ }\textbf {\bibinfo {volume} {58}},\ \bibinfo {pages}
  {949} (\bibinfo {year} {2014})}\BibitemShut {NoStop}%
\bibitem [{\citenamefont {Singh}\ and\ \citenamefont
  {Nott}(2003)}]{singh2003experimental}%
  \BibitemOpen
  \bibfield  {author} {\bibinfo {author} {\bibfnamefont {Anugrah}\ \bibnamefont
  {Singh}}\ and\ \bibinfo {author} {\bibfnamefont {Prabhu~R}\ \bibnamefont
  {Nott}},\ }\bibfield  {title} {\enquote {\bibinfo {title} {Experimental
  measurements of the normal stresses in sheared stokesian suspensions},}\
  }\href@noop {} {\bibfield  {journal} {\bibinfo  {journal} {J. Fluid Mech.}\
  }\textbf {\bibinfo {volume} {490}},\ \bibinfo {pages} {293--320} (\bibinfo
  {year} {2003})}\BibitemShut {NoStop}%
\end{thebibliography}%

\clearpage

\appendix
\section{Macroscopic Friction Coefficients}
\label{appendixa}
The detailed
form of $\mu(\sigma_{xy},\phi)$ has not been reported for shear thickening
systems although some work exists for viscous granular systems, which
permanently occupy the high-$\sigma$ shear-thickened state \cite{guy2015}.
Imposed-pressure measurements on frictional non-Brownian spheres
\cite{Boyer2011} and 2D simulations of circular discs \cite{trulsson2012} found
$\mu_{yy}(\phi)$ to be a monotonically decreasing function of $\phi$, tending
to a non-zero value $\mu_c$ at the jamming volume fraction $\phi_m$, itself a
function of the particle friction coefficient $\mu_p$. Data for different
$\mu_p$, including $\mu_p=0$, collapse onto the same master curve in the 2D
simulations. Since varying $\sigma_{xy}$ essentially shifts the jamming volume
fraction $\phi_J(\sigma_{xy})$ between $\phi_0$ and $\phi_m$, this collapse
implies that $\mu=\mu(\phi)$ only. A $\sigma_{xy}$-dependence may exist in 3D,
but we assume that this is small.

\cite{deboeuf2009} and \cite{Garland2013} have measured the $\phi$-dependence of the particle normal
stress in the vorticity direction $\sigma_{zz}$, but did not report
$\mu_{zz}(\phi)$. We can obtain $\mu_{zz}$ indirectly via the second normal
stress difference, $N_2$:
\begin{equation}
\label{eq:N2}
\frac{N_2}{\sigma_{xy}}=\frac{1}{\mu_{yy}}-\frac{1}{\mu_{zz}}.
\end{equation}
For shear thickening dispersions, $N_2$ is typically small in magnitude and
scales approximately with the shear stress for both continuous
\cite{Cwalina2014} and discontinuous \cite{Laun1994} thickening, implying that
the ratio $\mu_{zz}/\mu_{yy}$ is independent of $\sigma_{xy}$. In granular
suspensions,  $N_2/\sigma_{xy}$ has been found to vary only weakly with $\phi$
close to $\phi_m$ \cite{singh2003experimental}. Together these
observations suggest that, in the range of $\phi$ we are considering,
$\mu_{zz}$ is proportional to $\mu_{yy}$ and thus also slowly varying and
monotonic.

When plotting \autoref{fig:fc}(d) we took empirical expressions for
$\mu_{yy}(\phi, \phi_m)$ from \cite{Boyer2011} up to $\phi_m$, using our
value of $\phi_m=0.56$ (solid line, \autoref{fig:fc}(b), inset); above $\phi_m$
we use the form in \autoref{fig:fc}(c) (dashed line), which is a plausible
extension of the curve given results from 2-d simulations \cite{trulsson2012}.
The curves for $\sigma_{zz}$ (not shown) are qualitatively similar.
\vfil

\section{Supplementary argument for prohibition of vorticity bands}
\label{appendixb}

Separate equality of $p_s$ and $p_{zz}$ between bands implies that for steady-state vorticity banding
\begin{equation}
\label{eq:stability}
\sigma^{(1)}_{xy}/\mu_{zz}^{(1)}(\phi_1,f(\sigma^{(1)}_{xy})) = \sigma_{xy}^{(2)}/\mu_{zz}^{(2)}(\phi_2,f(\sigma^{(2)}_{xy})).
\end{equation}
Suppose first that $\phi_1 = \phi_2$. The impossibility of \autoref{eq:stability}
being obeyed is then easily seen by thinking about the special case of $\sigma$-independent friction depicted in \autoref{fig:fc}(d). With vorticity bands a vertical line segment must be found connecting two different points on the same blue curve (common $\gdot$ and $\phi$ but unequal shear stress). But this implies the existence of a similar line segment on the corresponding curve for $p_{zz}$ (which closely resembles the red curve shown for $p_{yy}$) so that that the relevant normal stress is also unequal. 

A little thought shows the same to hold generically even when $\mu_{zz}$ depends on stress via $f(\sigma)$, so long as this dependence is reasonable, such as the expected smooth evolution between two order-unity limits as $f$ varies from 0 to 1 \cite{Wyart2014}. Although exceptions might be created by fine-tuning the stress dependence of $\mu_{zz}$ in an exotic way, the generic physics is as follows. Steady vorticity bands are precluded because they need to be at the same particle pressure; but if they were, their frictional state and hence shear state would also be the same, leaving no difference between the bands. 

Vorticity bands with unequal concentration, $\phi_1\neq\phi_2$, can be excluded by a slight generalization of the same approach. Such bands require us to construct a vertical line connecting two blue curves such that the corresponding red curves are coincident at the chosen $\gdot$. If $\mu$ is a slowly-varying function of $\phi$ then no two red curves ever coincide except at the origin (see \autoref{fig:guilhem}(d)). If $\mu_{zz}(\phi)$ is strongly decreasing close to $\phi_m$ then one could construct a situation in which a high-$\sigma_{xy}$, low-$\phi$ phase coexists with a low-$\sigma_{xy}$, high-$\phi$ phase. (The converse situation arises when $\mu_{zz}(\phi)$ increases rapidly close to $\phi_m$.) But in that case, the ratio of $\mu_{zz}$ at $\phi_1$ and $\phi_2<\phi_1$ must be comparable to the ratio of the viscosities of the limiting quasi-Newtonian regimes at $\phi_2$:
\begin{equation}
\label{eq:jump}
\frac{\mu_{zz}(\phi_2)}{\mu_{zz}(\phi_1)} \sim \frac{\eta(\sigma_{xy}\ll\sigma^\star,\phi_2)}{\eta(\sigma_{xy}
\gg \sigma^\star, \phi_2)}
\end{equation}
For the parameters used to generate the flow curves in figures \ref{fig:fc} this requires $\mu_{zz}(\phi)$ to jump by a factor of $\mu_{zz}(0.553)/\mu_{zz}(0.558) \sim 10^2$ over a $\phi$-range of $0.005$. In the data of \cite{Boyer2011}, \autoref{fig:fc}(b), the change in $\mu_{yy}(\phi)$ is at most 10\% over the same range. By this argument, even allowing for particle migration, steady-state vorticity bands are physically precluded by equality of $p_{zz}$.

\section{Supplementary argument for prohibition of gradient bands}
\label{appendixc}

Gradient bands coexist at a common shear stress $\sigma_{xy}^{(1)}=\sigma_{xy}^{(2)}$ but different shear rates $\gdot_1 \neq \gdot_2$. The shear-thickening flow curves of interest have multi-valued $\sigma(\gdot)$ but single-valued $\gdot(\sigma)$. Crucially, this requires gradient bands always to have different concentrations, $\phi_1 \neq \phi_2$.  

Mechanical stability now demands that the normal stress component in the velocity gradient direction is continuous across the band interface, $\sigma_{yy}^{(1)}=\sigma_{yy}^{(2)}$. 
Using the same arguments as before to rule out spatial variations in solvent pressure $p_s$, we find the condition 
\begin{equation}
\label{eq:grad}
\mu_{yy}(\phi_1,f(\sigma_{xy}^{(1)}))=\mu_{yy}(\phi_2,f(\sigma_{xy}^{(2)}))=\mu_{yy}(\phi_2,f(\sigma_{xy}^{(1)})), 
\end{equation}
where the last equality follows from the common shear stress in the two bands. Graphically, in reference to \autoref{fig:fc}(d), steady gradient bands require us to find a horizontal line that connects two flow curves at different $\phi$ (blue lines) such that the corresponding $\sigma_{yy}$ values (red lines) are also equal. 
The latter is true if $\mu_{yy}$ is independent of $\phi$ (as was assumed for simplicity by WC and in \autoref{fig:fc}(d)) but is otherwise ruled out
for monotonic but non-constant $\mu_{yy}(\phi)$ of the kind generically expected in practice (compare \autoref{fig:fc}(c)). 

A possible exception again arises for the coexistence of a fully jammed state ($\phi>\phi_m,\gdot = 0$) with a flowing one ($\phi<\phi_m, \gdot > 0$). This outcome was reported by \cite{Fall2015}; however these authors used a wide-gap Couette system. In this geometry ``banding" is expected even for fixed-friction materials because the imposed ratio of shear to normal stress varies with radius. The interface between static and flowing ``bands" is where this ratio crosses the static friction threshold set by the repose angle in the material.

Assuming a constant ratio $\sigma_{xy}/\sigma_{yy} = \mu_J$ within the jammed band, then our argument still holds so long $\mu_J - \lim \mu_{yy}(\phi\to\phi_m^-)$ is either zero (as expected by continuity arguments), or has the same sign as $d\mu_{yy}/d\phi$ (in effect, maintaining monotonicity). However if $\mu_J$ is not constant but depends on other variables in the jammed state (such as a prior transient flow history, or an elastic strain) gradient banding is not necessarily ruled out. Yet it would require the dense, frictional, jammed band to maintain as low a normal stress as a more dilute, less frictional, flowing one of equal $\sigma_{xy}$. As discussed above for the case of vorticity bands, this reverses the usual expectation concerning the relative dilatancy and/or friction of these two types of packing.


\end{document}